\documentclass[a4paper,11pt]{article}

\usepackage{jheppub} 
\usepackage{amsmath}
\usepackage{graphicx,tabularx}
\usepackage{url}
\usepackage{footmisc}
\usepackage{amsfonts}
\usepackage{cancel}
\usepackage{color}
\usepackage{multirow} 
\usepackage{amssymb}
\usepackage{pifont}
\usepackage{epstopdf}
\usepackage{slashed}
\usepackage{comment}
\usepackage{booktabs} 
\usepackage{natbib}
\usepackage{array}
\usepackage{mathrsfs}
\usepackage{subcaption}
\usepackage[toc,page]{appendix}
\usepackage{mathtools}
\usepackage{romannum}
\usepackage{ulem}
\usepackage{bbold}
\allowdisplaybreaks

\def\lag{\mathscr{L}}

\def\beq{\begin{equation}}
\def\eeq{\end{equation}}

\title{Dark Matter Blind Spots at One-Loop}
\author[a,b,c]{Tao Han,}
\author[a]{Hongkai Liu,}
\author[d]{Satyanarayan Mukhopadhyay,}
\author[a]{Xing Wang}
\affiliation[a]{Department of Physics and Astronomy, University of Pittsburgh, Pittsburgh, PA 15260, USA }
\affiliation[b]{Department of Physics, Tsinghua University, Beijing, 100086, China}
\affiliation[c]{Collaborative Innovation Center of Quantum Matter, Beijing, 100086, China}
\affiliation[d]{School of Physical Sciences, Indian Association for the Cultivation of Science, Kolkata 700032, India}
\emailAdd{than@pitt.edu}
\emailAdd{hol42@pitt.edu}
\emailAdd{tpsnm@iacs.res.in}
\emailAdd{xiw77@pitt.edu}

\preprint{
\begin{flushright}
PITT-PACC-1815
\end{flushright}
}

\abstract{We evaluate the impact of one-loop electroweak corrections to the spin-independent dark matter (DM) scattering cross-section with nucleons ($\sigma_{\rm SI}$), in models with a so-called blind spot for direct detection, where the leading-order prediction for the relevant DM coupling to the Higgs boson, and therefore $\sigma_{\rm SI}$, are vanishingly small. 
Adopting a simple illustrative scenario in which the DM state results from the mixing of  electroweak singlet and doublet fermions, we compute the relevant higher order corrections to the scalar effective operator contributions to $\sigma_{\rm SI}$, stemming from both triangle and box diagrams involving the SM and dark sector fields. It is observed that in a significant region of the singlet-doublet model-space, the one-loop corrections ``unblind'' the tree-level blind spots and lead to detectable SI scattering rates at future multi-ton scale liquid Xenon experiments, with $\sigma_{\rm SI}$ reaching values up to a few times $10^{-47} {~\rm cm}^2$, for a weak scale DM with $\mathcal{O}(1)$ Yukawa couplings. Furthermore, we find that there always exists a new SI blind spot at the next-to-leading order, which is perturbatively shifted from the leading order one in the singlet-doublet mass parameters. For comparison, we also present the tree-level spin-dependent scattering cross-sections near the SI blind-spot region, that could lead to a larger signal. Our results can be mapped to the blind-spot scenario for bino-Higgsino DM in the MSSM, with other sfermions, the heavier Higgs boson, and the wino decoupled.}

\begin{document}

\titlepage

\maketitle

\newpage


\flushbottom

\section{Introduction}

Weakly interacting massive particles (WIMP), a possible candidate for the dark matter (DM) in the Universe, are being intensely searched for both in laboratory experiments and through a broad range of astrophysical probes~\cite{Jungman:1995df,Bertone:2004pz}. Among the laboratory probes, the decades-long programme looking for signals of nuclear recoil is the primary one, with increasing levels of sensitivity to the DM-nucleon scattering rate, owing to both larger fiducial detector volumes, as well as the construction of ultra-low noise detectors~\cite{Aprile:2018dbl,Akerib:2016vxi,Cui:2017nnn}. The current level of experimental sensitivity therefore calls for increased accuracy of the theoretical predictions as well, in order to thoroughly probe interesting and well-motivated WIMP scenarios. This becomes especially important if the leading order predictions for these scattering cross-sections are negligibly small or even exactly zero either due to symmetry reasons or due to cancellations among different contributions to the relevant DM effective couplings. Next-to-leading order corrections then become important, and would constitute a benchmark for the near-future multi-ton scale liquid Xenon-based direct detection experiments, targeting at a DM-nucleon scattering cross-section below $10^{-47} {\rm cm}^2$.

A well-studied example of the above scenario where the one-loop contributions to the DM-nucleon scattering rate become important is DM belonging to a multiplet of the Standard Model (SM) weak interaction group ${\rm SU} (2)_{\rm L}$~\cite{Cirelli:2005uq, Cirelli:2007xd}. For both real ${\rm SU} (2)_{\rm L}$ triplets with zero hypercharge (e.g., the wino in the minimal supersymmetric standard model, MSSM) and  Majorana ${\rm SU} (2)_{\rm L}$ doublets (e.g., the Higgsino in the MSSM) the leading contribution to spin-independent (SI) scattering with nucleons appears at one-loop. In the former case, the SI cross-section with nucleon is only mildly sensitive to the DM mass and is obtained to be around $2.3 \times 10^{-47} {\rm cm}^2$ in the limit $M_{\rm DM} \gg M_W$, including higher order corrections at next-to-leading order in $\alpha_s$~\cite{Hisano:2004pv, Hisano:2010fy, Hisano:2011cs, Hill:2013hoa, Hisano:2015rsa}. Therefore, these DM candidates are natural benchmark targets for multi-ton scale detectors. For Higgsino-like ${\rm SU}(2)_{\rm L}$ doublet Majorana fermions, the rate is further suppressed by two orders of magnitude, and the SI cross-section is around $10^{-49} {\rm cm}^2$. Such cross-sections are below the irreducible neutrino floor~\cite{Strigari:2009bq,  Billard:2013qya}, thereby making necessary larger detector volumes and exposure time, as well as the development of directional detection methods~\cite{Grothaus:2014hja, OHare:2015utx}. 

While for the pure ${\rm SU}(2)_{\rm L}$ multiplets discussed above the tree-level SI scattering rates are absent due to symmetry reasons, there are other scenarios in which very small tree-level rates are obtained due to cancellations of different contributions to the relevant effective couplings. For example, if the neutral components of different ${\rm SU} (2)_{\rm L}$ multiplets mix after electroweak symmetry breaking, generically there are regions of parameter space where the effective coupling to the Higgs boson(s), which determines the leading contribution to the SI scattering rate, either becomes small or even vanishes, a scenario dubbed as ``blind spots" for DM direct detection~\cite{Cheung:2012qy, Cheung:2013dua, Feng:2013fyw, Dedes:2014hga, Huang:2014xua, Crivellin:2015bva, Freitas:2015hsa, Badziak:2015exr, Banerjee:2016hsk, Han:2016qtc, Baum:2017enm}. While the particular values and relations of the theory parameters that result in the blind spots may not have any deeper theoretical implications, or may even be viewed as a fine-tuning to a special hypersurface within the parameter space, they do characterize a distinctive class of phenomena that need to be scrutinized. Such blind spots for DM-nucleon scattering therefore present us with another context in which the higher-order electroweak corrections, involving states from both the dark matter and the SM sectors in the loop amplitudes, are important to evaluate in order to quantify its detectability. In this paper, we compute the one-loop corrections to DM-nucleon scattering processes near such blind spots, and assess their implications for different direct detection probes.

As an example scenario, which represents all the features of more involved models such as the bino-Higgsino mixed DM in the MSSM~\cite{Jungman:1995df}, we begin by studying a DM model with mixing between an ${\rm SU} (2)_{\rm L} \times {\rm U}(1)_{\rm Y}$ singlet fermion and the neutral components of two ${\rm SU} (2)_{\rm L}$ doublet fermions~\cite{Mahbubani:2005pt, DEramo:2007anh,Banerjee:2016hsk}. The details of this simplified model and the appearance of tree-level blind spots are reviewed in Section~\ref{sec:SD}. We then systematically evaluate the impact of the one-loop corrections for the SI scattering rates near the blind spots
in the singlet-doublet model, after defining an on-shell renormalization procedure for the dark matter sector. The computational framework and the results of the one-loop corrections are discussed in Section~\ref{sec:NLRC}, while the details of the on-shell renormalization scheme adopted are summarized in Appendix~\ref{renormalization}. In Section~\ref{sec:cons} we utilize these one-loop results to find out the prospects of observing DM-nucleon scattering near the tree-level blind spots. In this section, we also compare the prospects for probing the one-loop SI rates with the reach from the tree-level spin-dependent (SD) DM-nucleon scattering searches. We provide a summary of our study in Section~\ref{sec:sum}. We also briefly review the computational framework adopted in this paper for SI and SD DM-nucleon scattering in Appendix~\ref{DD}, and the mapping of the singlet-doublet model parameters to the case for MSSM bino-Higgsino mixed DM scenario in Appendix~\ref{SUSY}.


\section{Singlet-doublet dark matter and tree-level blind spot}
\label{sec:SD}

To understand the appearance of blind spots for DM direct detection, it is instructive to consider a simple model, in which the DM candidate is a linear combination of an electroweak singlet Majorana fermion $\chi_S$, and the neutral components of two ${\rm SU} (2)_{\rm L}$ doublet states $\chi_{D1}$ and $\chi_{D2}$, with hypercharge $+1/2$ and $-1/2$, respectively~\cite{Cheung:2012qy, Cheung:2013dua,Banerjee:2016hsk},
\beq
\chi_{D1}=(\chi_1^+,\chi_1^0)^\top \quad\text{and}\quad \chi_{D2}=(\chi_2^0,\chi_2^-)^\top.
\eeq
The mixing between the singlet and the neutral components of the doublet states occurs after electroweak symmetry breaking. Such a scenario can appear in beyond-the-standard-model constructions such as the MSSM, in which the singlet state is the bino, and the two doublet states correspond to the two Higgsinos. In the MSSM some of the couplings of these states with the SM sector are determined by gauge symmetry and supersymmetry, and therefore the results of the singlet-doublet model can be mapped to the MSSM case, as long as all the sfermions, heavy scalars and wino are decoupled.

In order to have a stable DM candidate, we impose an additional $Z_2$ symmetry, under which the DM sector states are odd, and all the SM sector states are even. Thus, the lightest neutral state in the dark sector is the DM candidate, where the mass spectrum and Yukawa couplings of the dark sector particles are determined by the following Lagrangian
\beq
\lag_Y=- \left(\frac{1}{2}M_S\chi_S \chi_S+ M_D\chi_{D1}\cdot\chi_{D2}-y_1\chi_S\chi_{D1}\cdot\widetilde{H}-y_2\chi_S\chi_{D2}\cdot H \right)+ {\rm h.c.},
\label{eq:lag_mass}
\eeq
where $H= \left(\phi^+,(v+h+i\eta)/\sqrt{2} \right)^\top$ is the SM Higgs doublet, with a vacuum expectation value $v = 246$ GeV, while $\widetilde{H}=i\sigma_2H^*$. The dot products in Eq.~(\ref{eq:lag_mass}) indicate the contraction of ${\rm SU} (2)_{\rm L}$ indices to form a singlet.

We see that the mass spectrum is determined by four free parameters, namely, $M_D,~M_S,~y_1$ and $y_2$. By re-defining the fields $\chi_{D1}$, $\chi_{D2}$, and $\chi_S$, we can make three of them positive, chosen to be $y_1$, $y_2$, and $M_S$.
For simplicity, we do not include any possible CP violation in the DM sector, and restrict to real values of $M_D$ only. After electroweak symmetry breaking, the neutral components of the doublet and singlet dark fermions mix, and the mass matrix of neutral dark sector in the gauge basis $\chi^0=(\chi_S,-\chi_2^0,\chi_1^0)^\top$ is given by
\beq
\mathbf{M_N}=\begin{pmatrix}
	M_S & {yv\cos\beta \over \sqrt{2}} & {yv\sin\beta \over \sqrt{2}} \\
	\frac{yv\cos\beta }{\sqrt{2}} & 0 &  M_D\\
	\frac{yv\sin\beta }{\sqrt{2}} &  M_D & 0\\
\end{pmatrix},
\eeq
where $\tan\beta = y_1/y_2$, with $y_1=y\sin\beta$ and $y_2=y\cos\beta$.

\subsection{Spin-independent interaction}
The dominant contribution to SI direct detection cross-section stems from the Higgs boson exchange diagram, and we obtain the tree-level DM-Higgs coupling using the low energy theorem
\beq
C^{0}_{h\widetilde{\chi}^0_1\widetilde{\chi}^0_1}=\frac{1}{2}\frac{\partial M_{\widetilde{\chi}^0_1}(v)}{\partial v}
=\frac{y^2 v[M_D\sin(2\beta)+M_{\widetilde{\chi}^0_1}]}{6M_{\widetilde{\chi}^0_1}^2-4M_{\widetilde{\chi}^0_1}M_S-2M^2_D-y^2v^2} \ ,
\label{coup}
\eeq
where $\widetilde{\chi}^0_1$ is the lightest neutral mass eigenstate.
In the limit of vanishing momentum transfer relevant for nuclear-recoil experiments, the SI direct detection rate is fixed by the Wilson coefficient $f_q$ of the operator $m_q\overline{\widetilde{\chi}^0_1}\widetilde{\chi}^0_1\overline{q}q$. The $t$-channel Higgs exchange process leads to the following isospin-conserving Wilson coefficient for interactions with up-type and down-type quarks
\beq
f_u=f_d=-\frac{C^{0}_{h\widetilde{\chi}^0_1\widetilde{\chi}^0_1}}{vm_h^2} \ .
\eeq
There is an additional effective coupling to a pair of gluons in the nucleon, which is obtained on integrating out the heavy quarks coupled to the Higgs propagator, and the corresponding Wilson coefficient is given as \cite{Jungman:1995df}
\beq
f_G=-\frac{1}{12}\sum_{q=c,b,t}f_q \ .
\eeq
Combining the quark and the gluon contributions, we obtain the effective coupling of the DM state to nucleons, $f_N \overline{\widetilde{\chi}^0_1}\widetilde{\chi}^0_1\overline{N}N$, with
\beq
\begin{split}
	f_N/m_N =\sum_{q=u,d,s}f_qf_{Tq}^N+\frac{2}{27}\sum_{q=c,b,t}f_qf^N_{TG} 
	 =-\frac{C^{0}_{h\widetilde{\chi}^0_1\widetilde{\chi}^0_1}}{vm_h^2}\left[\sum_{q=u,d,s}f^N_{Tq}+\frac{2}{9}f^N_{TG}\right],
\label{FNTree}
\end{split}
\eeq
where $f_{Tq}^N$ and $f^N_{TG}$ are the mass-fraction parameters of the quarks and the gluon in the nucleon $N$, respectively, and $f^N_{TG}\equiv 1-\sum_{q=u,d,s}f^N_{Tq}$. We have summarized the additional details in the computation of DM-nucleon scattering in Appendix~\ref{DD}.

Thus, we see from the above discussion that at the leading order, the SI DM-nucleon scattering rate via the Higgs boson exchange would vanish if the mass and Yukawa coupling parameters satisfy the following blind-spot condition~\cite{Cheung:2012qy, Cheung:2013dua, Banerjee:2016hsk}
\beq
M_D\sin(2\beta)+M_{\widetilde{\chi}_1^0}=0.
\label{BDcond}
\eeq
For our choice of the phases of the mass and Yukawa coupling parameters, we see that the blind-spot condition can be satisfied for $M_D<0$. For the specific choice of parameters that satisfy the blind-spot condition, since the coupling of the DM mass eigenstate to the Higgs boson is zero, the physical mass of the DM state is either $M_S$ or $M_D$, depending upon the hierarchy. Thus the two possibilities are
\begin{enumerate}
	\item $M_{\widetilde{\chi}_1^0} = M_S$, $-M_D > M_S $,\ \  $\sin(2\beta)=M_S/(-M_D)$,
	\item $M_{\widetilde{\chi}_1^0} =- M_D$, $-M_D < \left(M_S+\sqrt{M_S^2+(yv)^2} \right)/2 $,\ \  $\tan\beta=1$.
\end{enumerate}
While the first possibility leads to an SI blind spot, the second one implies a blind spot for both SI and SD scattering. For our subsequent analyses, we take up the first case as an illustration.

\subsection{Spin-dependent interaction}
In the singlet-doublet model, the spin-dependent interaction of DM with the nucleon is determined by the gauge interaction of the doublet components with the $Z$-boson. The relevant interaction Lagrangian is given in terms of the gauge eigenstates by
\beq
\begin{split}
	\lag_{int} = - {e \over 2 \cos\theta_W \sin\theta_W}  \left[(\chi_1^0)^\dagger \sigma_-^\mu\chi_1^0 -  (\chi_2^0)^\dagger \sigma_-^\mu\chi_2^0  \right ]Z_\mu , 
\end{split}
\label{EWcoupling}
\eeq
where $\theta_W$ is the Weinberg angle. Thus the axial-vector coupling of the DM state to the $Z$-boson, which leads to the spin-dependent interaction with nucleons is obtained to be
\beq
C^{0}_{Z\widetilde{\chi}^0_1\widetilde{\chi}^0_1}=\frac{e}{2 s_W^{} c_W^{}}(U_{21}^2-U_{31}^2),
\eeq
where $s_W=\sin\theta_W,\ c_W=\cos\theta_W$,  and the mixing matrix in the neutral dark sector is defined by
\beq
\widetilde{\chi}^0=U^\dagger \chi^0,
\eeq
with the mass eigenstates $\widetilde{\chi}^0=\left({\widetilde{\chi}^0_1},{\widetilde{\chi}^0_2},\widetilde{\chi}^0_3 \right)^\top$. Therefore, the Wilson coefficient of the relevant low-energy effective interaction $\overline{\widetilde{\chi}^0_1}\gamma^\mu\gamma^5\widetilde{\chi}^0_1\overline{q}\gamma_\mu\gamma^5q$ is found to be (please see Appendix~\ref{DD} for further details on the standard formalism adopted) 
\beq
d_u=\frac{-e^2(U_{21}^2-U_{31}^2)}{8 M_Z^2 s_W^2 c_W^2} = -d_d \ .
\eeq

\section{Radiative corrections to DM-nucleon scattering}
\label{sec:NLRC}

We now turn to the electroweak radiative corrections to the spin-independent DM direct detection rate near the tree-level blind spots. Since the SI scattering rates are vanishingly small around this region of mass and coupling parameters, the next-to-leading order (NLO) corrections are expected to play an important role in determining the detectability of such DM model-space. Furthermore, as we will see in the following, there also appears a new blind spot at NLO order, at a shifted parameter region compared to the tree-level one.

\subsection{Computational framework}
In addition to the interaction Lagrangians described in Eqs.~(\ref{eq:lag_mass}) and (\ref{EWcoupling}), the following additional interaction terms (in the gauge basis) involving the charged components of the DM doublets and the weak bosons enter the computation of the radiative corrections
\beq
\begin{split}
	\lag_{int} =& \frac{e}{\sqrt{2}\ s^{}_W} \left[ \left( (\chi_1^0)^\dagger \sigma_-^\mu\chi_1^+ + (\chi_2^-)^\dagger \sigma_-^\mu\chi_2^0 \right)W_\mu^- + {~\rm h.c.}\right].
\end{split}
\label{EWcoupling2}
\eeq

\begin{figure}[h!]
	\begin{subfigure}{.5\textwidth}
		\centering
		\includegraphics[width=0.5\textwidth]{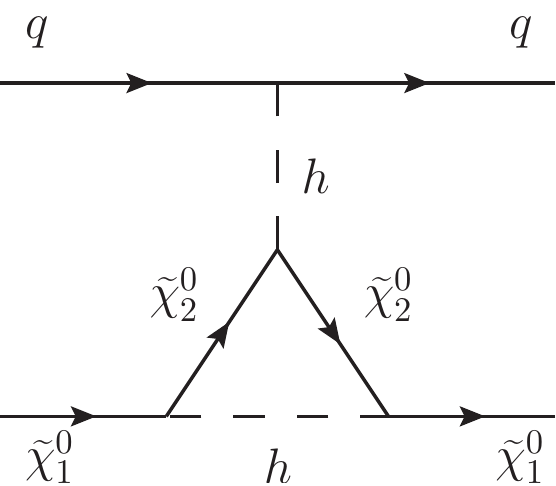}
		\caption{}
		\label{tri}
	\end{subfigure}
	\begin{subfigure}{.5\textwidth}
	\centering
	\includegraphics[width=0.5\textwidth]{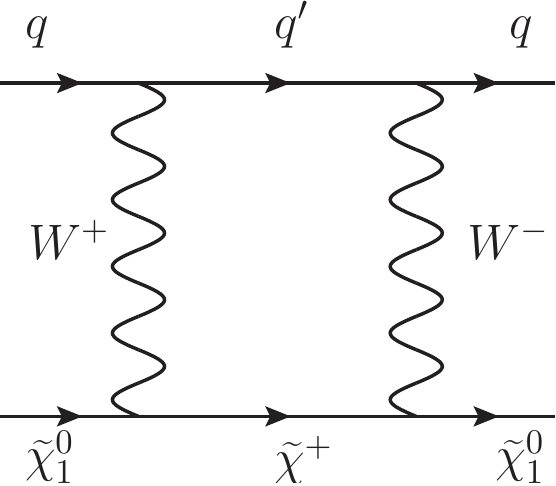}
	\caption{}
	\label{box}
\end{subfigure}
	\caption{Representative Feynman diagrams contributing to DM-quark spin-independent scattering.}
\label{fig:oneloop}
\end{figure}

There are two different amplitudes contributing to the NLO electroweak corrections to DM-nucleon scattering, with representative Feynman diagrams depicted in Fig.~\ref{fig:oneloop}. The first one stems from the one-loop vertex corrections to the Higgs-DM coupling, as shown in Fig.~\ref{tri}, while the second one is given by the box diagrams shown in Fig.~\ref{box}. Since the triangle diagrams are ultraviolet (UV) divergent, we need to renormalize the relevant mass, mixing and coupling parameters. We have adopted the on-shell renormalization scheme for the dark matter sector, the details of which are described in Appendix~\ref{renormalization}. 

In addition to the class of diagrams represented in Fig.~\ref{fig:oneloop}, there are other sets of diagrams entering the NLO electroweak corrections to the same process. These involve the Higgs self-energy corrections and the vertex corrections to the quark Yukawa couplings. However, the contribution of these latter diagrams to the DM-quark effective vertex is proportional to the tree-level DM-Higgs coupling, which is vanishingly small near the tree-level blind-spot region of our interest. We therefore focus on the diagrams in Fig.~\ref{fig:oneloop} for our computation, which constitute a UV-finite subset.

We have generated the relevant Feynman diagrams and the corresponding matrix elements using {\it FeynArts}~\cite{Hahn:2000kx}, which are then passed onto {\it FeynCalc}~\cite{Shtabovenko:2016sxi,Mertig:1990an} to perform the Passarino-Veltman reduction of the one-loop integrals. We have used Collier~\cite{Denner:2016kdg,Denner:2002ii,Denner:2005nn,Denner:2010tr} for the numerical evaluation of the one-loop scalar integrals. We have adopted the Feynman gauge for our computations.


\begin{figure}[t]
	\begin{subfigure}{.5\textwidth}
		\centering
		\includegraphics[width=\textwidth]{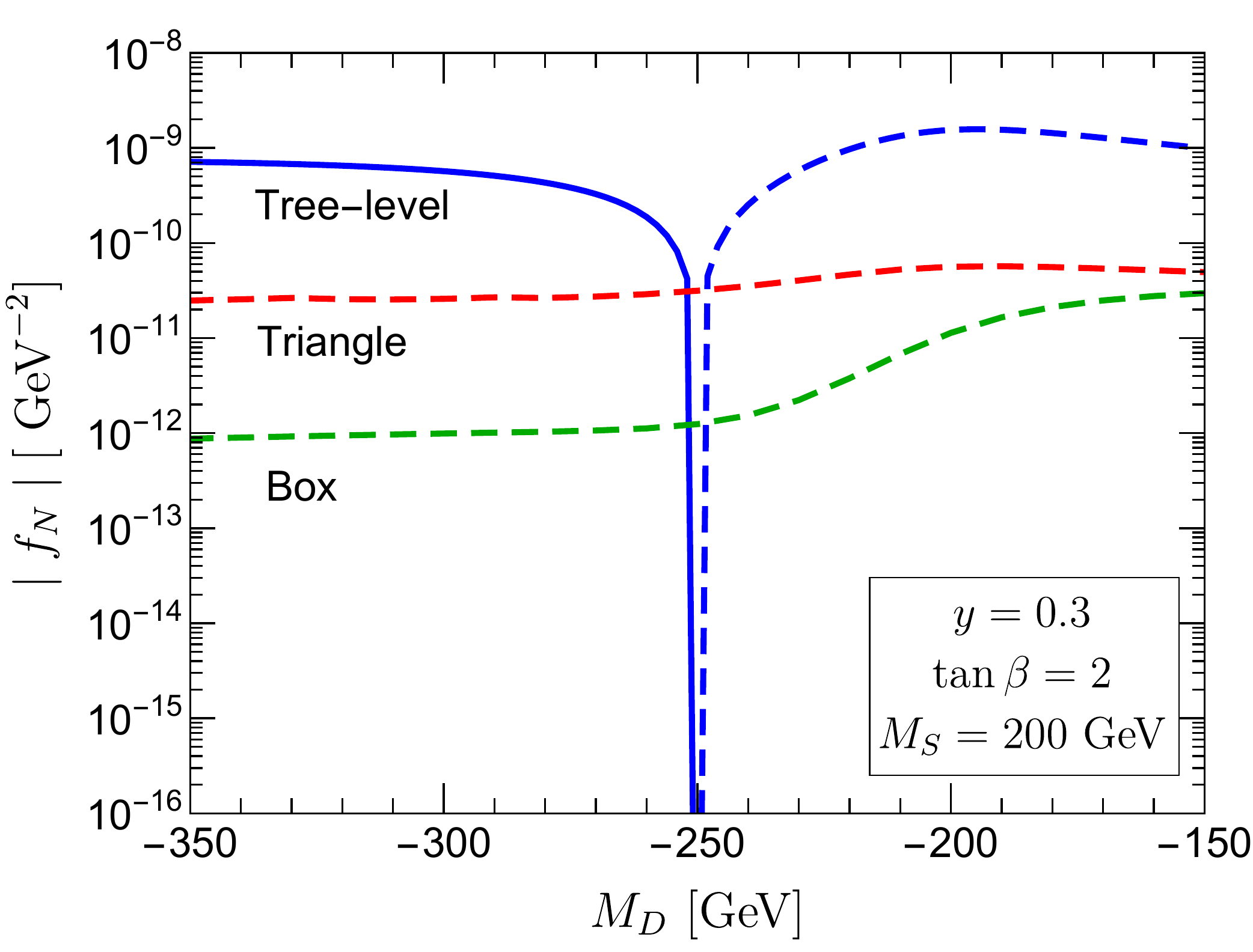}
		\caption{}
		\label{amp1}
	\end{subfigure}~~~
	\begin{subfigure}{.5\textwidth}
		\centering
		\includegraphics[width=\textwidth]{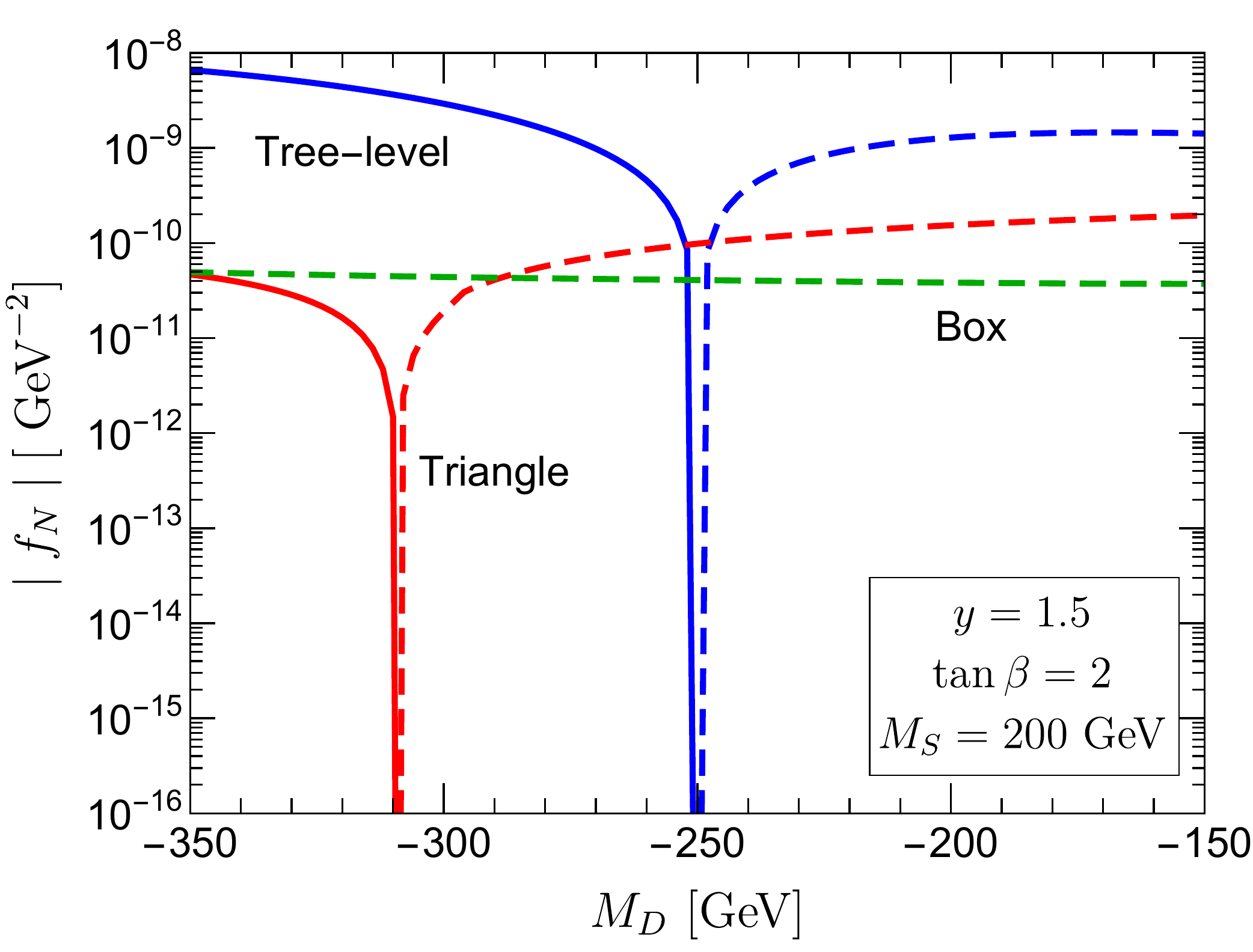}
		\caption{}
		\label{amp2}
	\end{subfigure}
		\begin{subfigure}{.5\textwidth}
		\centering
		\includegraphics[width=\textwidth]{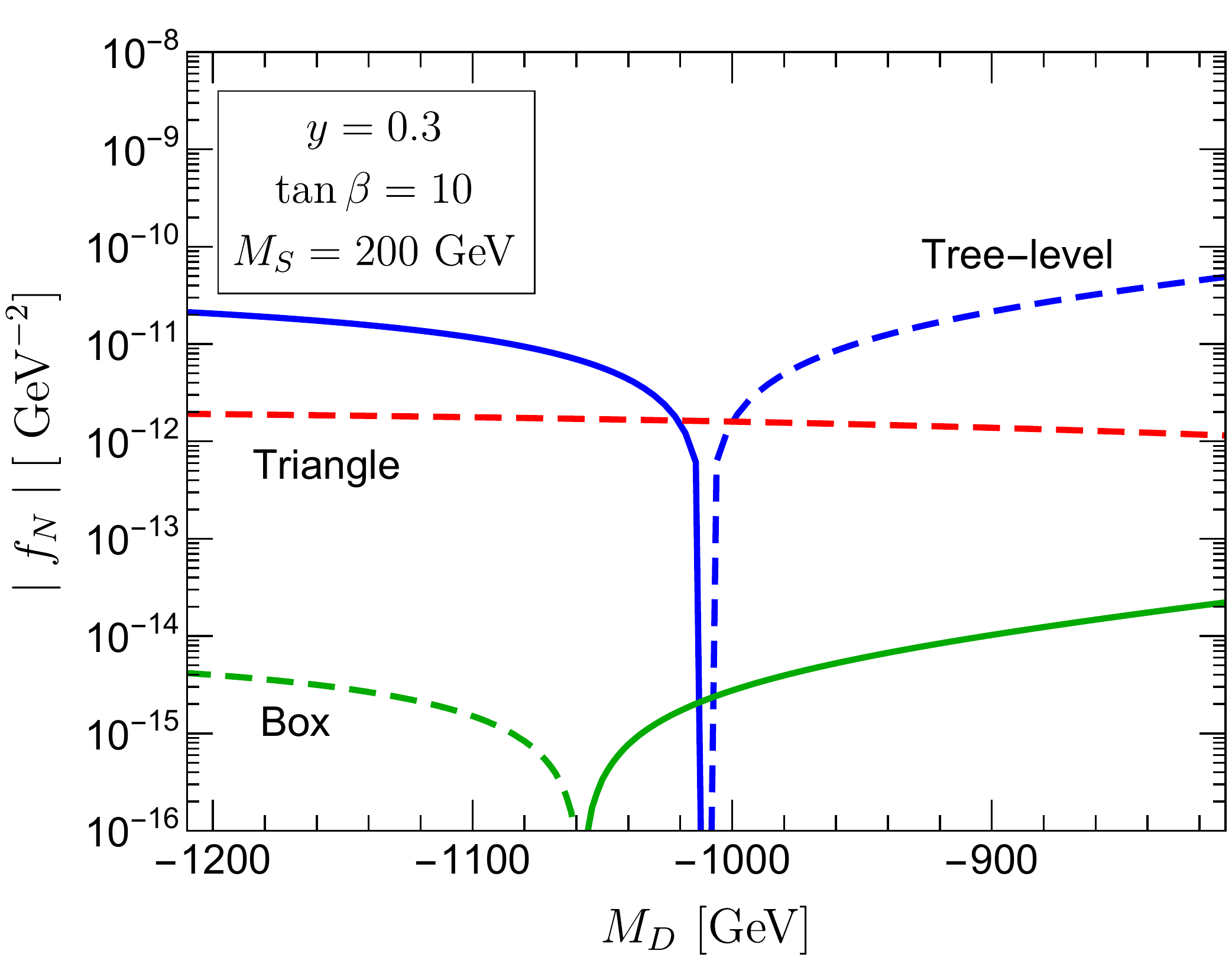}
		\caption{}
		\label{amp3}
	\end{subfigure}~~~~
	\begin{subfigure}{.5\textwidth}
		\centering
		\includegraphics[width=\textwidth]{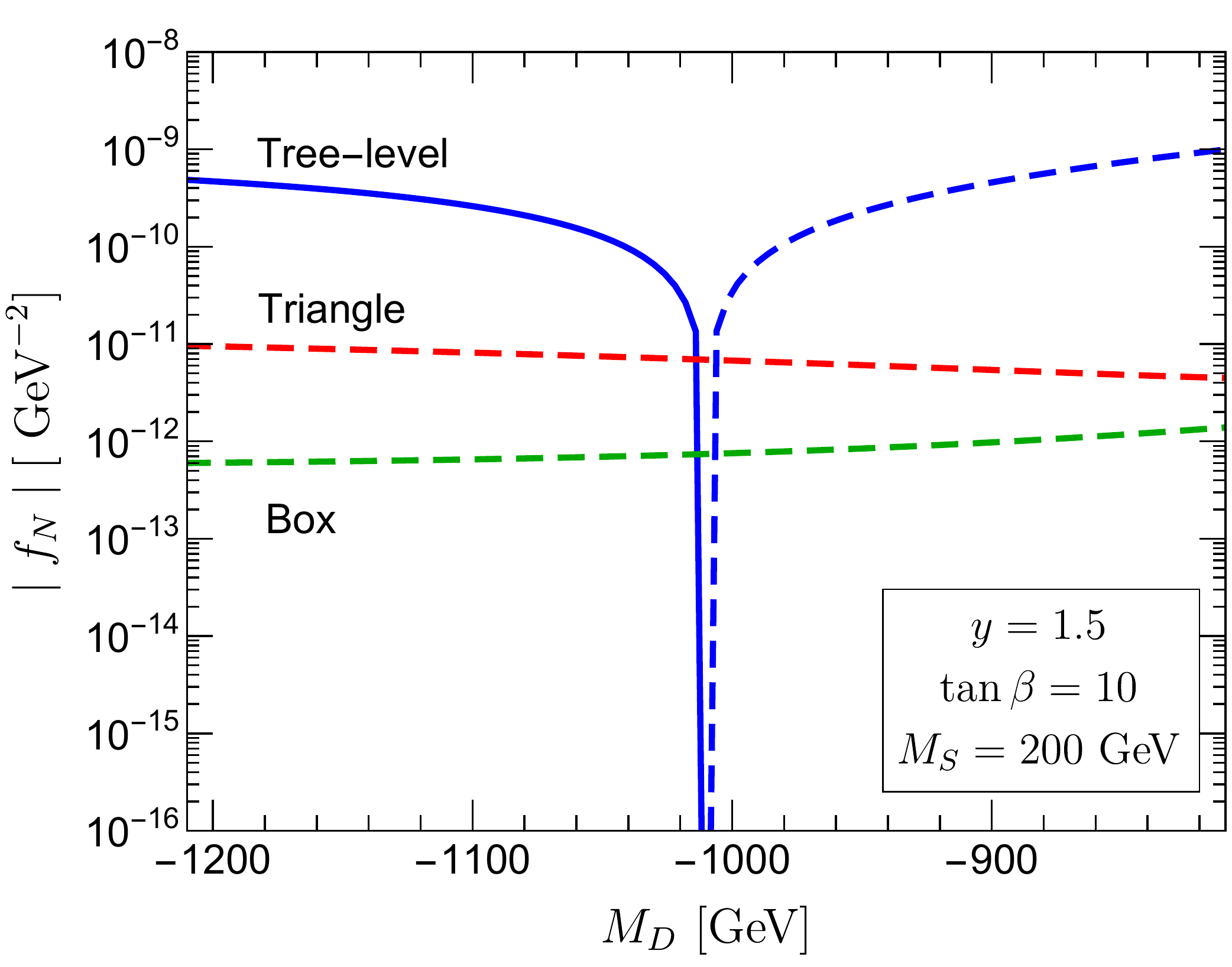}
		\caption{}
		\label{amp4}
	\end{subfigure}

	\caption{Contributions to the absolute value of $f_N$ as a function of $M_D$, from tree-level diagrams (blue), one-loop triangle diagrams (red), and one-loop box diagrams (green). The dashed lines indicate negative values of $f_N$. The value of the singlet dark fermion mass is fixed as $M_S=200$ GeV, with $\tan \beta = 2$ (upper panels) and $\tan \beta = 10$ (lower panels), for representative values of $y=0.3$ (left columns) and $y=1.5$ (right columns).}
	\label{fig:amp}
\end{figure} 

\subsection{Results}

The contribution to the effective DM-quark interaction from the vertex corrections represented by the triangle diagrams in Fig.~\ref{tri}, $f_N^{\rm tri}$, has the same form as the tree-level $t$-channel Higgs exchange vertex, with the Higgs-DM coupling $C_{h\widetilde{\chi}^0_1\widetilde{\chi}^0_1}^0$ replaced by its one-loop counterpart $C_{h\widetilde{\chi}^0_1\widetilde{\chi}^0_1}^{\rm tri}$
\beq
f_N^{\rm tri}/M_N=-\frac{C_{h\widetilde{\chi}^0_1\widetilde{\chi}^0_1}^{\rm tri}}{vm_h^2}\left(\sum_{q=u,d,s}f_{Tq}+\frac{2}{9}f_{TG}\right).
\eeq
The box diagrams shown in Fig.~\ref{box} also induce corrections to the Wilson coefficient of the operator $\overline{\widetilde{\chi}^0_1}\widetilde{\chi}^0_1\overline{q}q$, denoted as $C_q^{\rm box}$, which are not universal for different flavors, and lead to the following corrections to the DM-nucleon effective scalar coupling:
\beq
f_N^{\rm box}/M_N=\sum_{q=u,d,s}\frac{C_q^{\rm box}}{m_q}f_{Tq}+\frac{2}{27}f_{TG}\sum_{q=c,b,t}\frac{C_q^{\rm box}}{m_q},
\label{eq:coef_box}
\eeq
The ``long-distance'' contribution~\cite{Hisano:2011cs, Hisano:2015rsa} induced by the loop-diagrams to DM-gluon interaction is given by the second term in Eq.~(\ref{eq:coef_box}).

We show the resulting magnitudes of the tree-level $f_N^{\rm tree}$, the triangle diagram $f_N^{\rm tri}$, and the box diagram $f_N^{\rm box}$ contributions as a function of $M_D$ in Fig.~\ref{fig:amp}, where we have adopted the Feynman gauge for our computations. The results are shown for $M_S=200$ GeV with various values of $y$ and $\tan \beta$. Here, dashed lines have been used to indicate negative values of the Wilson coefficients. We note several interesting features in Fig.~\ref{fig:amp}. First of all, although the tree-level contribution naturally dominates in the parameter region away from the blind spot, near the blind spot it decreases dramatically. The one-loop contribution, especially from the triangle diagrams, therefore gives rise to the leading contribution in this region. Secondly, away from the blind spot, the one-loop electroweak effects are still appreciable. For example, we see in Figs.~\ref{amp2} and \ref{amp3} that the contributions from the triangle diagrams considered can shift the tree-level results by up to $10\%$. Third, the box diagram contribution can be comparable to the triangles in certain regions of parameter space. Fourth, there are values of parameters around which the triangle and the box contributions can change sign individually, and therefore have their own blind spots, as seen in Figs.~\ref{amp2} and \ref{amp3}. 

\begin{figure}[th]
	\begin{subfigure}{.5\textwidth}
		\centering
		\includegraphics[width=0.97\textwidth]{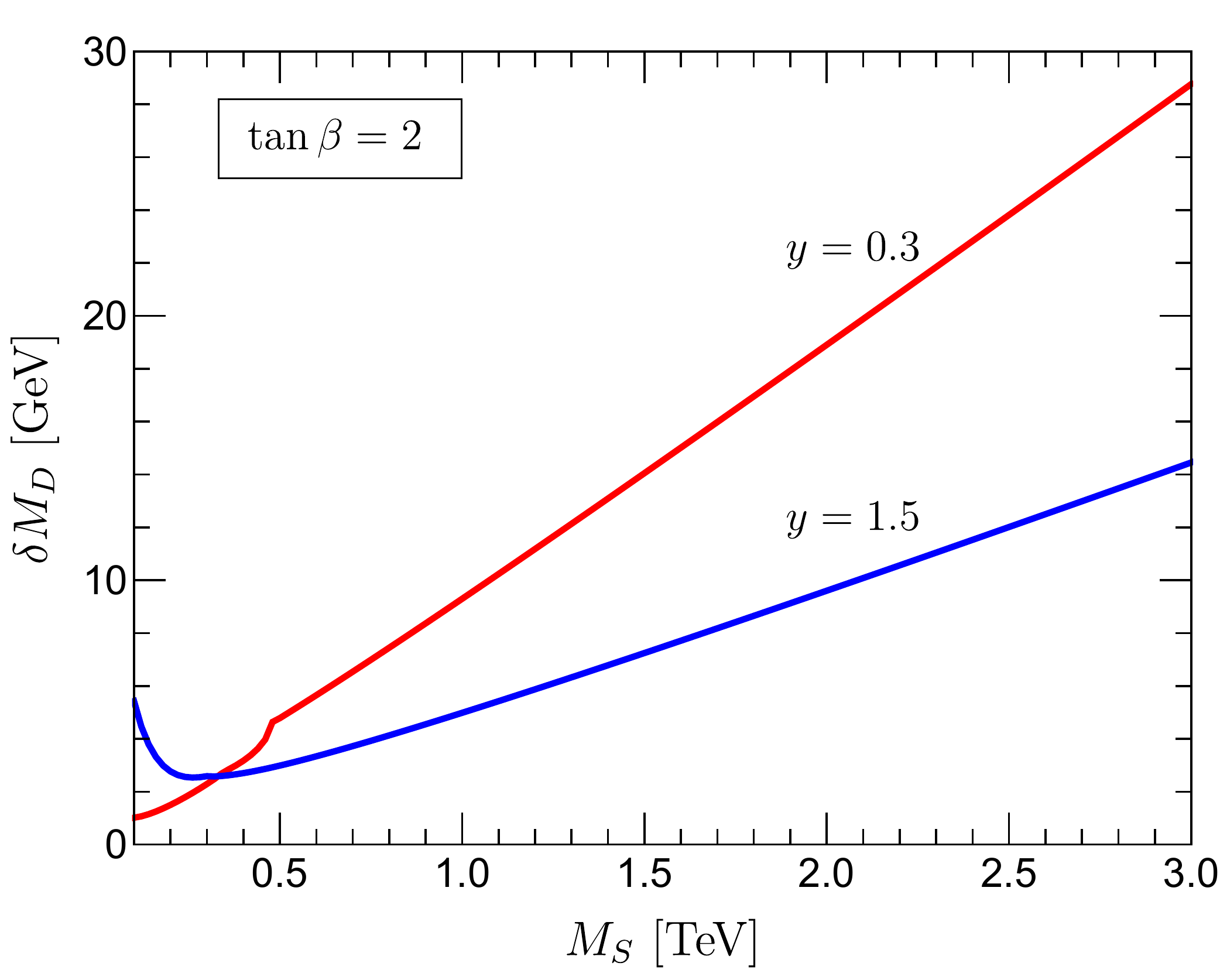}
		\caption{}
		\label{fig:DBD1}
	\end{subfigure}~~~%
	\begin{subfigure}{.5\textwidth}
		\centering
		\includegraphics[width=\textwidth]{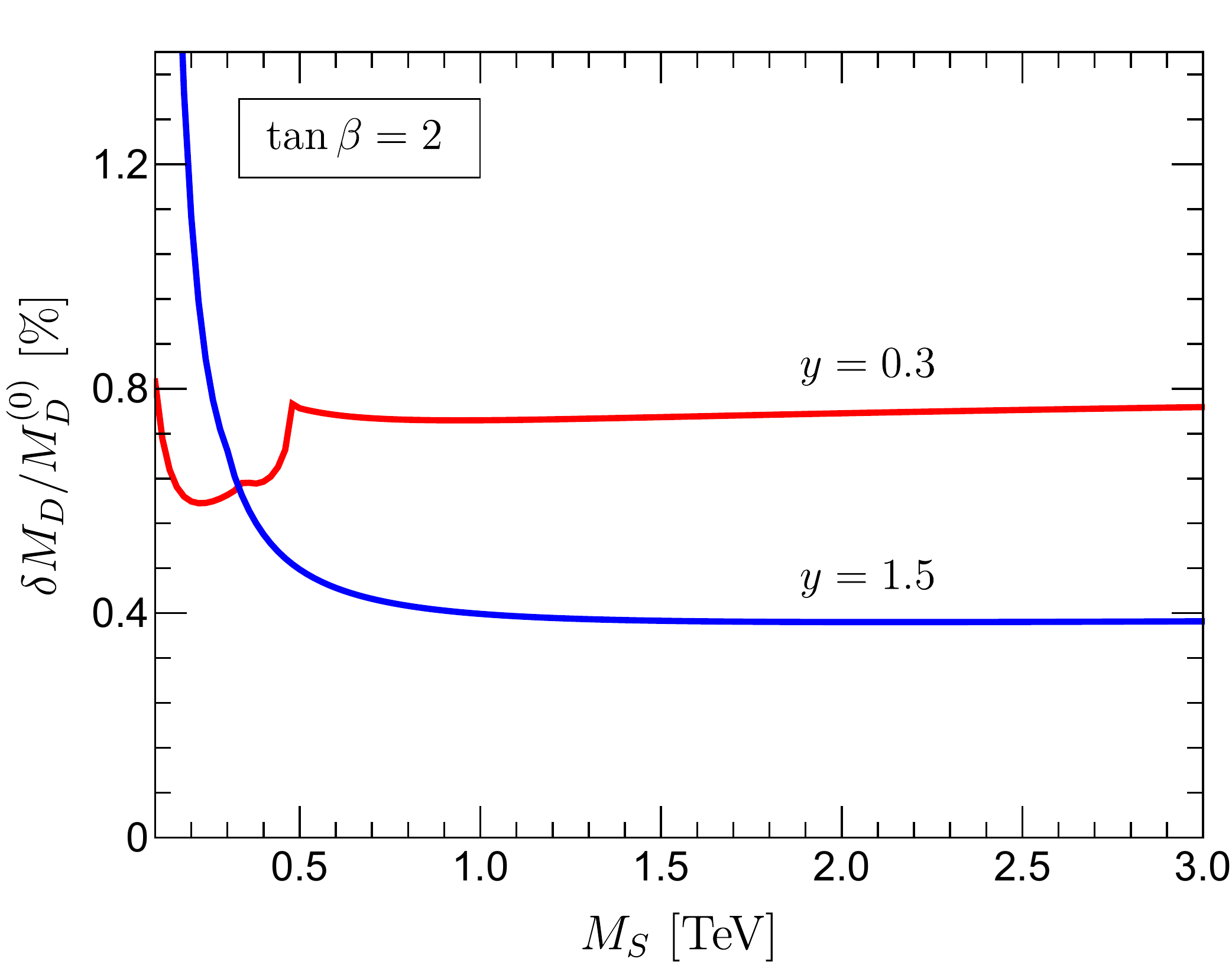}
		\caption{}
		\label{fig:DBD2}
	\end{subfigure}
	\begin{subfigure}{.5\textwidth}
		\centering
		\includegraphics[width=0.97\textwidth]{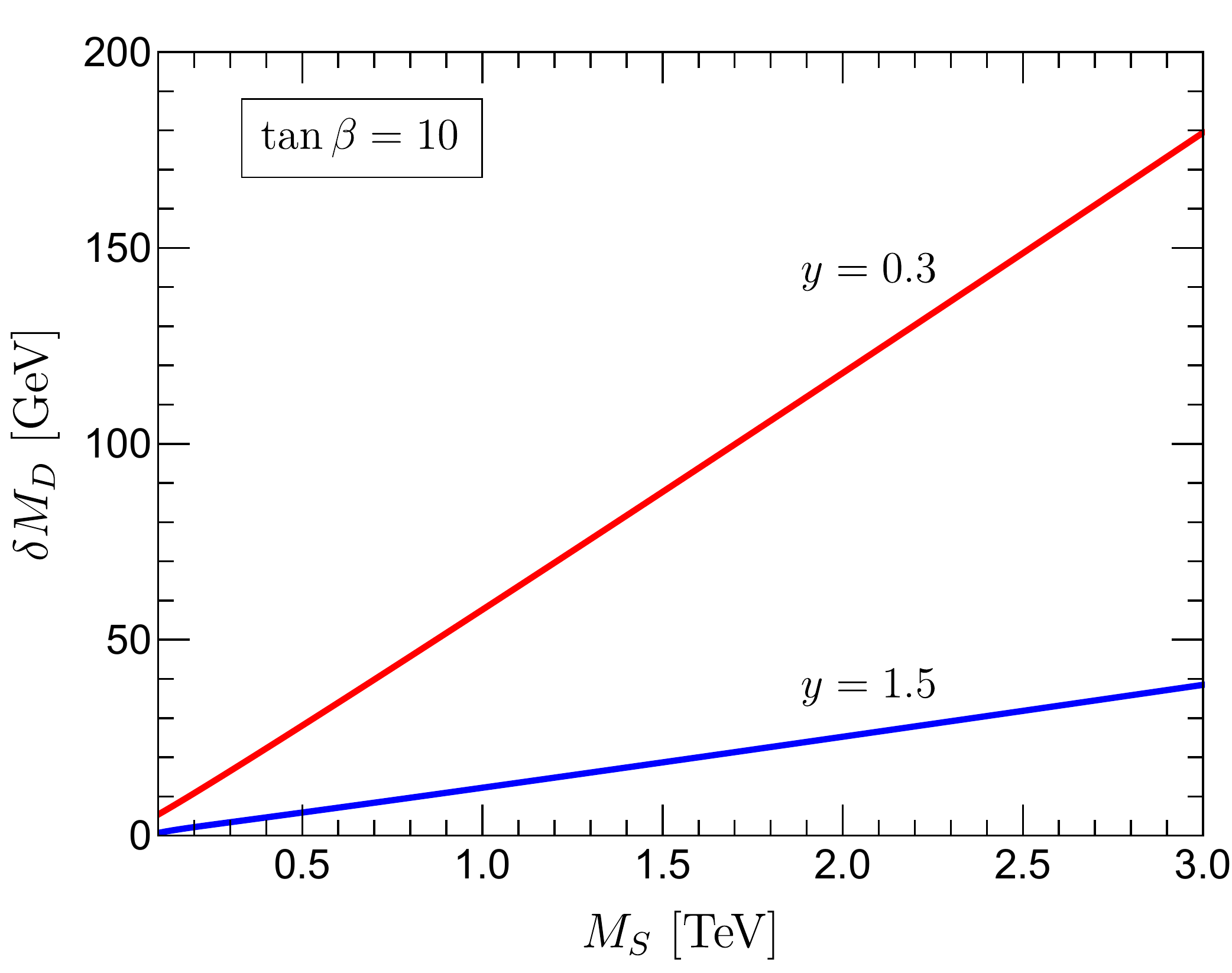}
		\caption{}
		\label{fig:DBD3}
	\end{subfigure}~~~%
	\begin{subfigure}{.5\textwidth}
		\centering 
		\includegraphics[width=\textwidth]{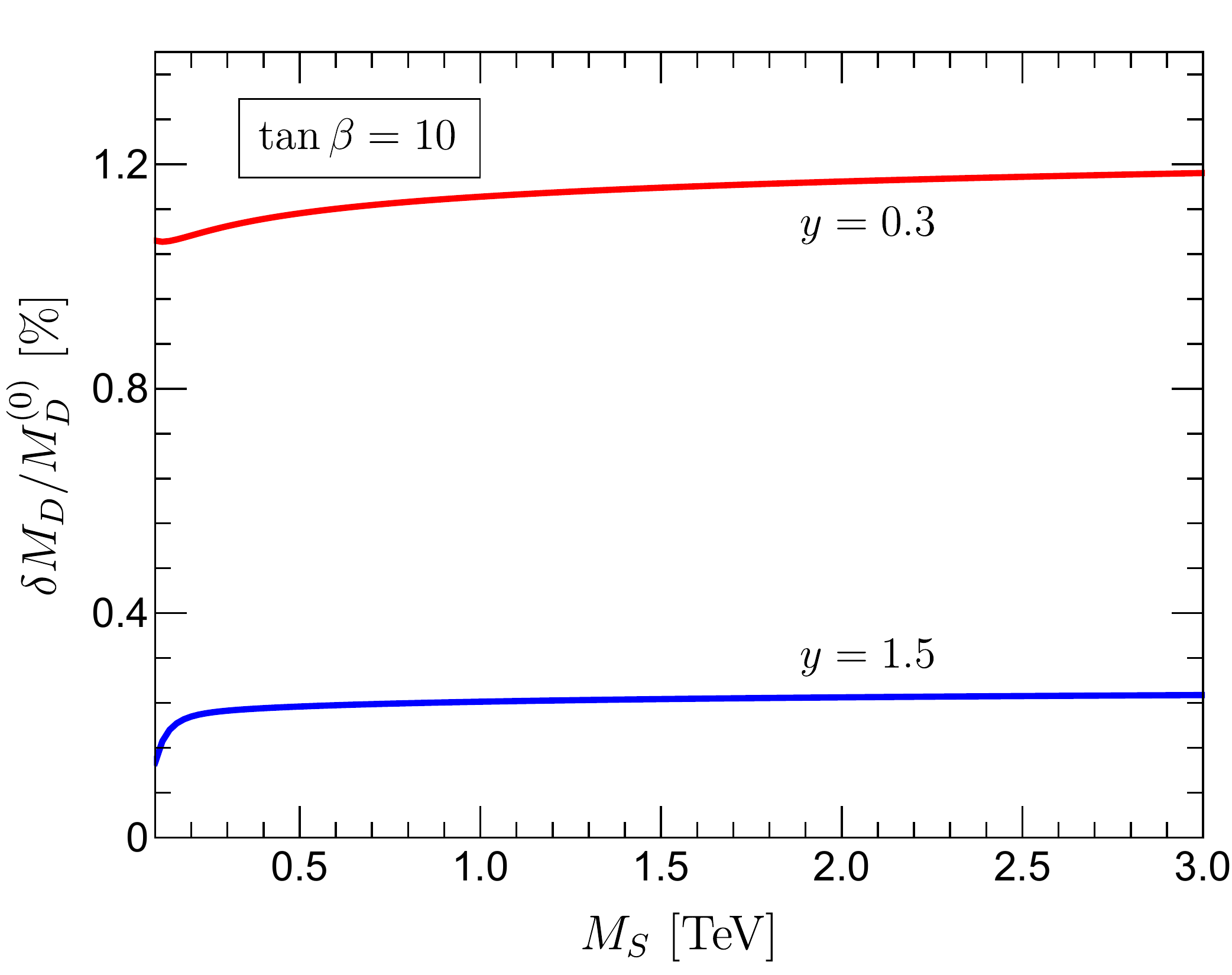}
		\caption{}
		\label{fig:DBD4}
	\end{subfigure}
	\caption{Shift in the position of the blind spot $\delta M_D$ versus $M_S$, with $\delta M_D=M_D^{(0)}-M_D^{(1)}$.
The results are shown for two values of the coupling $y=0.3$ (red) and $y=1.5$ (blue), with $\tan \beta=2$ (upper panels) and $\tan \beta=10$ (lower panels). We also show the ratio $\delta M_D/M_D^{(0)}$ in the right panels.}
	\label{fig:DBD}
\end{figure}

Most importantly, the full amplitude, which is a coherent sum of all the diagrams, always shows a new blind spot at the NLO level, perturbatively shifted from the tree-level blind spot. We quantify this shift by introducing a mass parameter difference 
\beq
\delta M_D=M_D^{(0)}-M_D^{(1)} ,
\eeq
which is the difference between the tree-level blind spot $M_D^{(0)} = -M_{\widetilde{\chi}^0_1}/\sin2\beta$ and the new blind spot $M_D^{(1)}$ obtained at NLO, on including the one-loop corrections. This variation in $\delta M_D$  is shown in Fig.~\ref{fig:DBD} as a function of $M_S$. The amount of the shift in the values of $M_D$ is  almost linearly proportional to the value of $M_S$ as seen in Figs.~\ref{fig:DBD1} and \ref{fig:DBD3}. The results are shown for two values of the coupling $y=0.3$ (red) and $y=1.5$ (blue), with $\tan\beta=2$ (upper panels) and $\tan\beta=10$ (lower panels). As we can see from this figure, the shift is larger for large values of $\tan \beta$ and small values of $y$. We also show the ratio $\delta M_D/M_D^{(0)}$ in Figs.~\ref{fig:DBD2} and \ref{fig:DBD4} (the right panels), and it can be around $\mathcal{O}(1\%)$ for small values of $y$. 

We note that the red curves in Figs.~\ref{fig:DBD1}~and~\ref{fig:DBD2}, with $y=0.3$ and $\tan\beta=2$, exhibit two cusps at $M_S \simeq 330$~GeV and $470$~GeV. These are due to the opening of new thresholds where the decays $\widetilde{\chi}^0_{2,3} \rightarrow \widetilde{\chi}^0_1\,Z$ and $\widetilde{\chi}^0_{2,3} \rightarrow \widetilde{\chi}^0_1\,h$, respectively, become kinematically accessible\footnote{The masses of $\widetilde{\chi}^0_2$ and $\widetilde{\chi}^0_3$ are nearly degenerate close to the tree-level blind spot parameter region.}. On the other hand, the blue curves in Figs.~\ref{fig:DBD1}~and~\ref{fig:DBD2}, and all the curves in Figs.~\ref{fig:DBD3}~and~\ref{fig:DBD4} do not have such cusps, as the decay channels $\widetilde{\chi}^0_{2,3} \rightarrow \widetilde{\chi}^0_1\,Z$ and $\widetilde{\chi}^0_{2,3} \rightarrow \widetilde{\chi}^0_1\,h$ are always allowed in the relevant parameter regions.


\section{Direct detection: current constraints and future prospects}
\label{sec:cons}

We now apply the results of the previous section to estimate the reach of ongoing and future direct detection experiments in the singlet-doublet model parameter space near the tree-level blind spot region. After discussing the NLO contribution to the spin-independent scattering, we also show the LO estimate for the reach of spin-dependent scattering experiments for comparison.

\subsection{Spin-independent scattering cross-sections at one-loop}
\label{subsec:SI}

In this section, we focus on the parameter region for the tree-level SI blind spot, where the NLO corrections are most impactful in extending the reach of SI direct detection probes. For a fixed value of $\tan \beta$, this then leads to a two-dimensional parameter space of interest, that of the DM mass ($M_{\widetilde{\chi}^0_1}$) and Yukawa coupling $y$ plane. The value of $M_D$, for each $M_{\widetilde{\chi}^0_1}$, is fixed to be $-M_{\widetilde{\chi}^0_1}/\sin \left(2\beta \right)$ as given by the blind-spot condition in Eq.~(\ref{BDcond}). 

\begin{figure}[t]
	\begin{subfigure}{.5\textwidth}
		\centering
		\includegraphics[width=\textwidth]{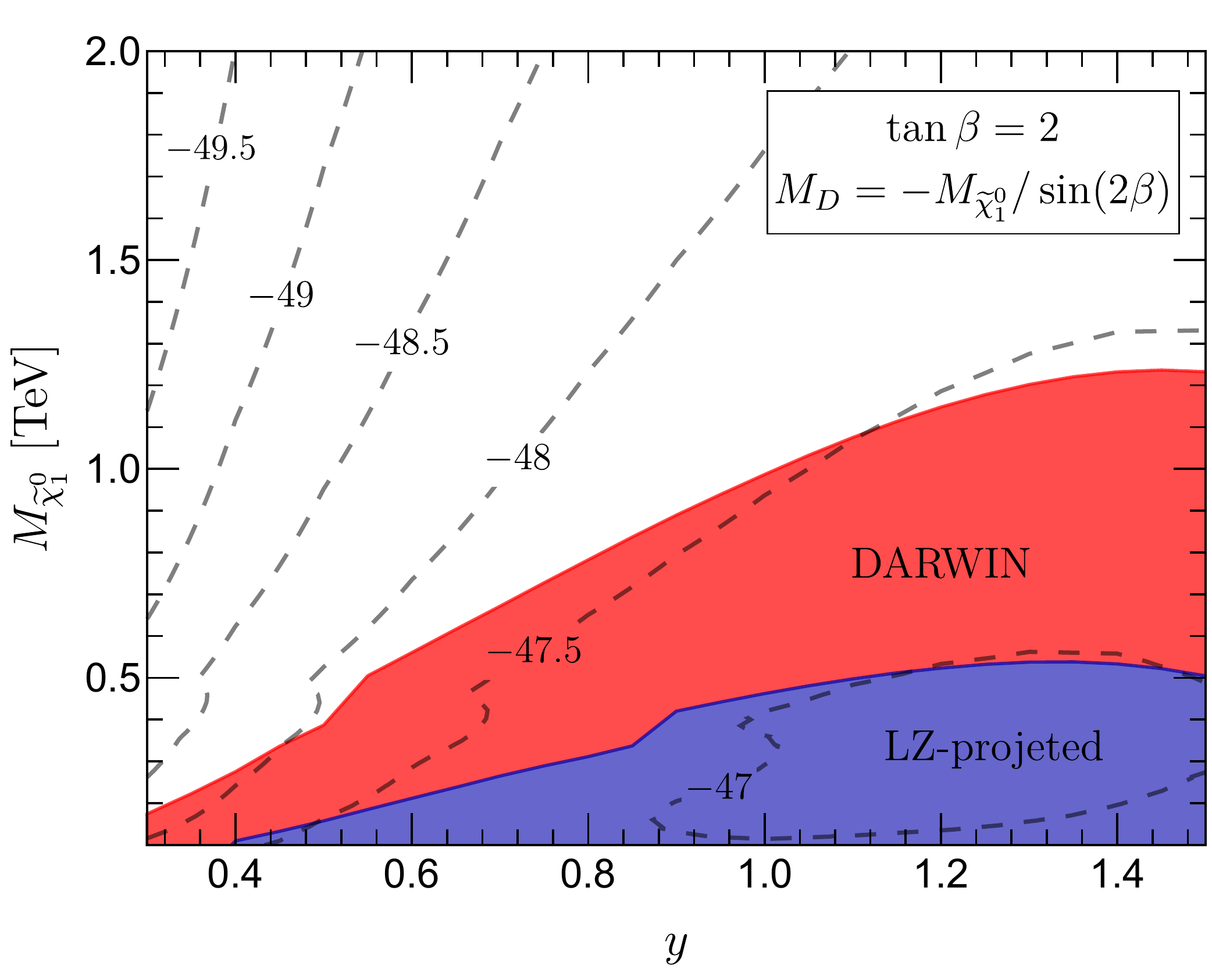}
		\caption{}
		\label{1BD1}
	\end{subfigure}~~~%
	\begin{subfigure}{.5\textwidth}
		\centering
		\includegraphics[width=\textwidth]{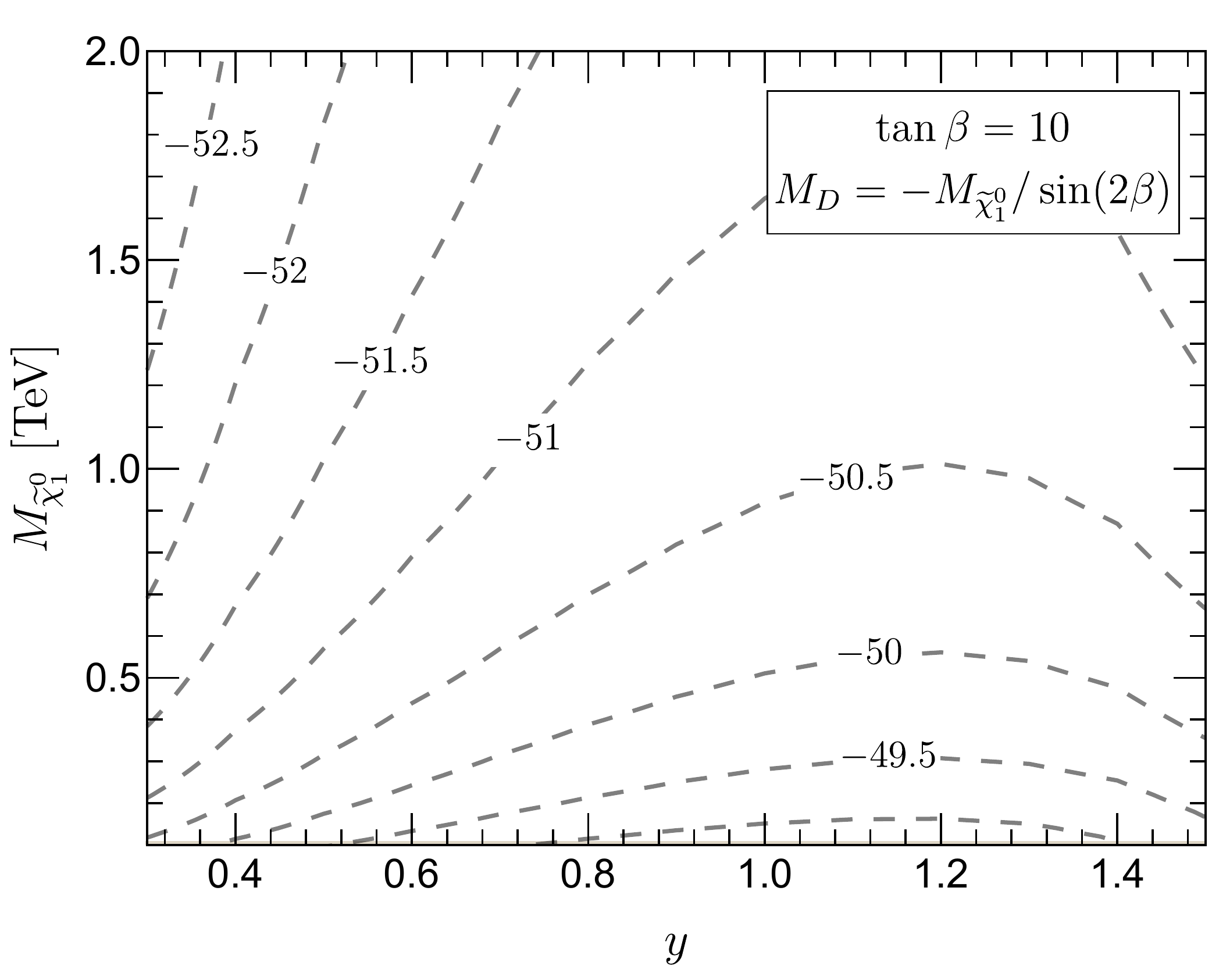}
		\caption{}
		\label{1BD2}
	\end{subfigure}%
	\caption{Spin-independent DM-nucleon scattering cross-section ($\sigma_{\rm SI}$) in the DM mass ($M_{\widetilde{\chi}^0_1}$)--Yukawa coupling ($y$) plane, with $M_D$ fixed by the blind-spot condition. The results are shown for $\tan\beta=2$ (left panel) and $\tan\beta=10$ (right panel). The contours represent lines with fixed values of $\log_{10}{\sigma^{\rm SI}}$, with $\sigma_{\rm SI}$ expressed in ${\rm cm}^2$ units. The projected reach of the LZ~\cite{Akerib:2018lyp}~(blue shaded) and DARWIN~\cite{Aalbers:2016jon}~(red shaded) experiments are also shown, with the DM candidate assumed to saturate the observed relic density.}
	\label{fig:SIBD}
\end{figure} 

In Fig.~\ref{fig:SIBD}, we show the contours of SI DM-nucleon scattering cross-section, $\sigma_{\rm SI}$,  in the $y-M_{\widetilde{\chi}^0_1}$ plane, for values of $\tan\beta=2$ (left) and $\tan\beta=10$ (right). 
As we can see, for $\tan \beta=2$, $\sigma_{\rm SI}$ takes values in the range of about $10^{-47} {~\rm cm}^2$ to $10^{-50} {~\rm cm}^2$, for $M_{\widetilde{\chi}^0_1}$ values in the interval $100 {~\rm GeV} - 2 {~\rm TeV}$, and coupling coefficient $y$ in the range $0.3-1.5$. For a given coupling, the cross-section decreases with increasing DM mass, and the future projection from the LZ experiment~\cite{Akerib:2018lyp}~(blue shaded region) is expected to probe a DM mass upto about $500$ GeV (blue shaded region in Fig.~\ref{fig:SIBD}), for the above range of $y$, assuming the DM candidate to saturate the observed relic density. This reach can be further extended by the DARWIN experiment~\cite{Aalbers:2016jon}~(red shaded region), which can probe DM masses of upto $1250$ GeV for the same range of coupling values. For higher values of $\tan \beta$, as seen with $\tan \beta=10$ in the right panel of Fig.~\ref{fig:SIBD}, the expected cross-section is smaller due to the suppression from smaller mixing angles, with a maximum of around $10^{-49} {~\rm cm}^2$, which may not be accessible to DARWIN. Thus, the small $\tan \beta$ scenario leads to similar $\sigma_{\rm SI}$ as in the case of wino-like real triplet DM, as discussed in the introduction, while the intermediate $\tan \beta$ scenario predicts cross-sections similar to the case of Majorana Higgsino-like doublets.

\subsection{Tree-level spin-dependent scattering cross-sections}
\label{subsec:SDXS}
In the spin-independent (SI) blind-spot region considered above, the effective coupling of the DM mass eigenstate to the Higgs boson vanishes. On the other hand, the spin-dependent (SD) scattering rate, which is determined at the tree level by the DM-$Z$-boson coupling, can have an appreciable rate for the same set of model parameters. In general, though the experimental sensitivity of SD scattering is weaker than that of SI scattering, near the blind spot they might have comparable reach~\cite{Cheung:2012qy, Calibbi:2015nha, Han:2016qtc}, since the SI rates appear only at NLO.

\begin{figure}[t]
	\begin{subfigure}{.5\textwidth}
		\centering
		\includegraphics[width=\textwidth]{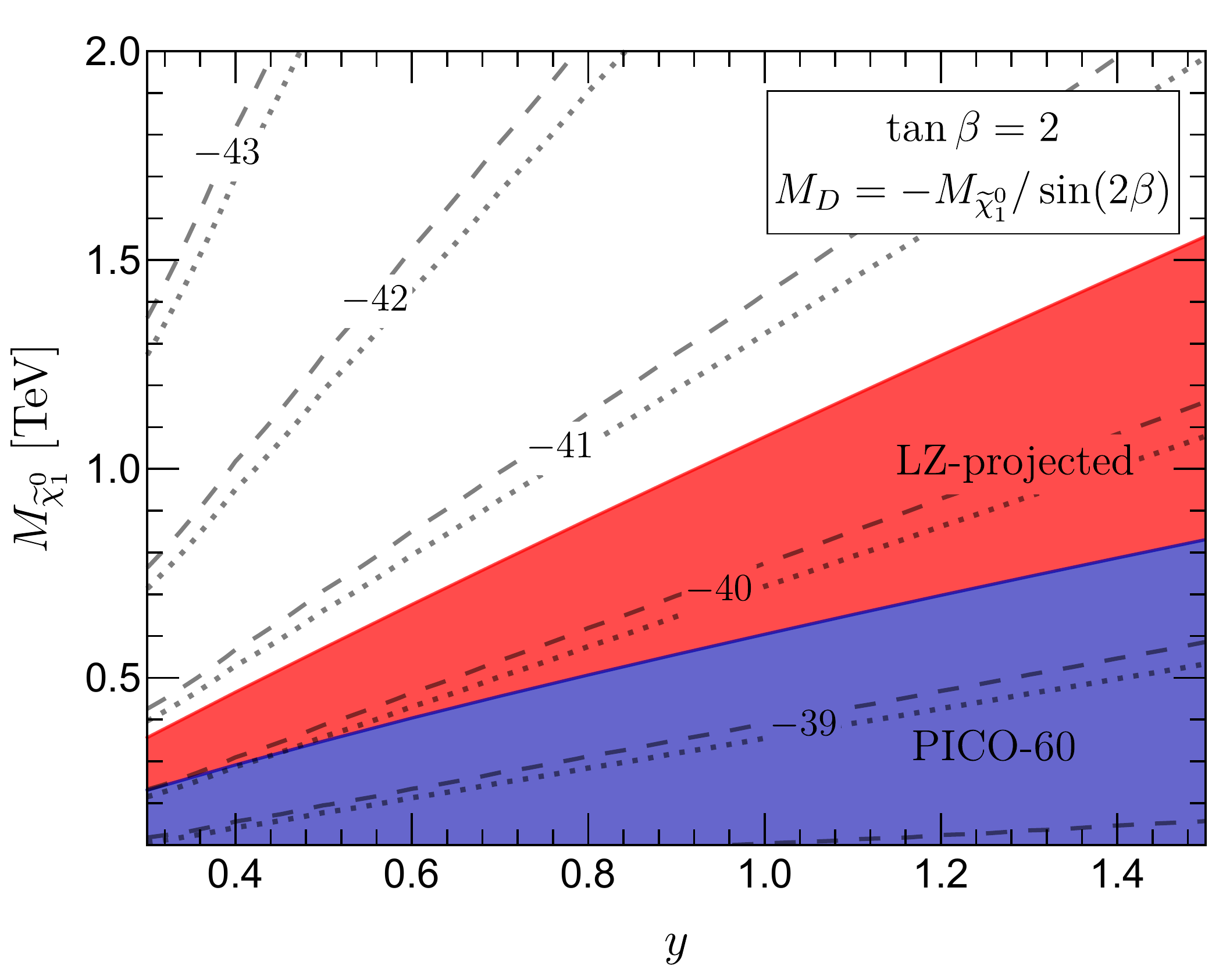}
		\caption{}
		\label{BDSD1}
	\end{subfigure}~~~
	\begin{subfigure}{.5\textwidth}
		\centering
		\includegraphics[width=\textwidth]{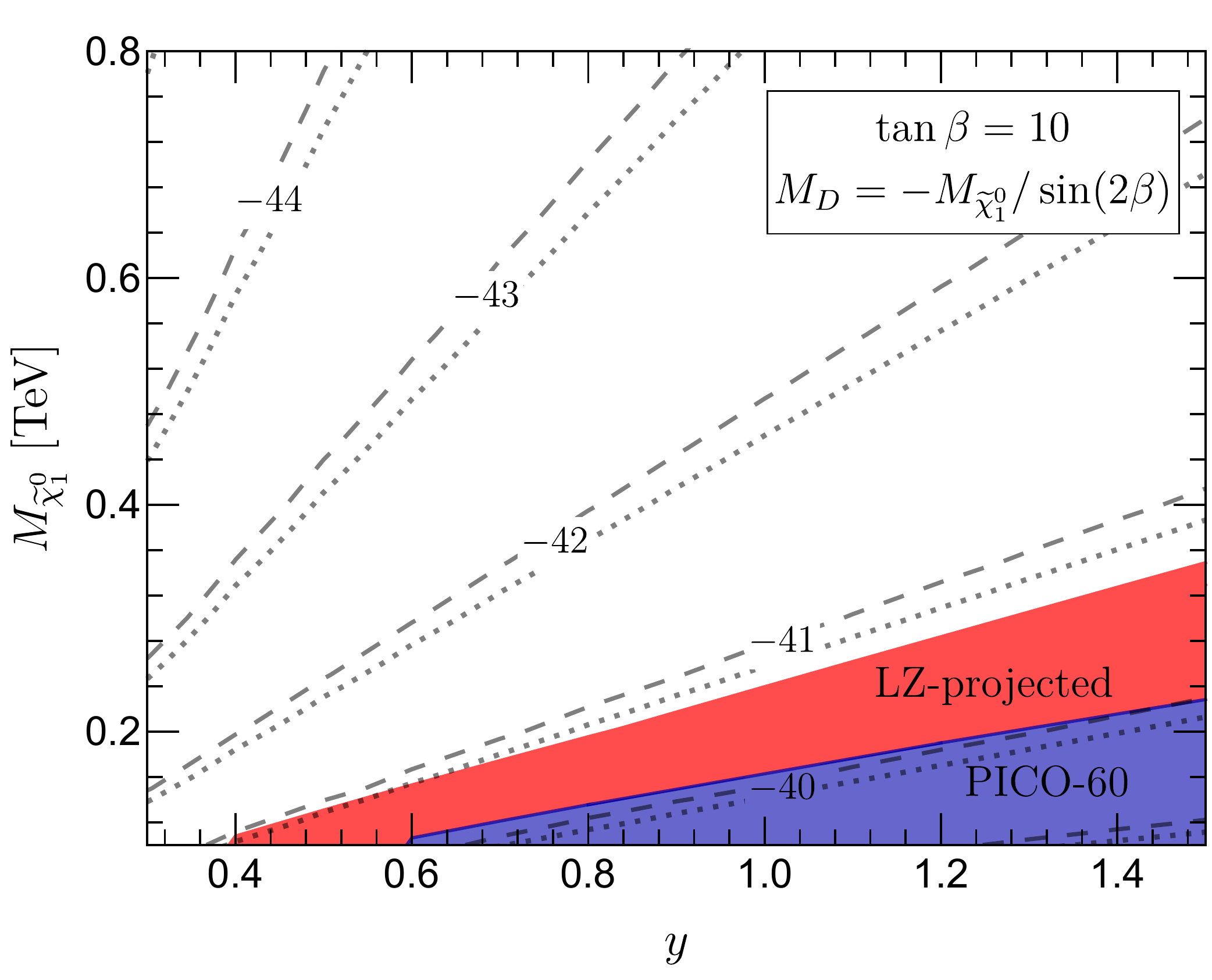}
		\caption{}
		\label{BDSD2}
	\end{subfigure}
	\caption{Spin-dependent DM-nucleon scattering cross-sections ($\sigma^{p,n}_{\rm SD}$) in the DM mass ($M_{\widetilde{\chi}^0_1}$)--Yukawa coupling ($y$) plane, with all other parameters and conditions being the same as in Fig.~\ref{fig:SIBD}. The contours represent lines with fixed values of $\log_{10}{\sigma^p_{\rm SD}}$ (dashed) and $\log_{10}{\sigma^n_{\rm SD}}$ (dotted), with $\sigma^{p,n}_{\rm SD}$ expressed in ${\rm cm}^2$ units. The reach of the ongoing PICO-60 experiment (blue shaded) and the projected reach of the LZ experiment (red shaded) are also shown.}
	\label{fig:SDlimit}
\end{figure}

We show the spin-dependent scattering cross-sections, $\sigma^{p}_{\rm SD}$ for proton and $\sigma^n_{\rm SD}$ for neutron, in Fig.~\ref{fig:SDlimit} in the $y-M_{\widetilde{\chi}^0_1}$ plane, with all other parameters and conditions  being the same as in Fig.~\ref{fig:SIBD}~\footnote {We note that, since the parametric shift of the blind spot from tree-level to NLO is only of $\mathcal{O}(1\%)$, the difference between SD cross-sections at the tree-level and NLO SI blind spots are negligible.}. The corresponding cross-sections are in the range of $10^{-38} - 10^{-43} {~\rm cm}^2$ for $\tan \beta=2$, and around an order of magnitude lower for $\tan \beta=10$, in the parameter space studied. The reach from the current PICO-60 experiment~\cite{Amole:2015lsj,Amole:2015pla,Amole:2017dex} (blue shaded region) and the future projections from the LZ experiment~~\cite{Akerib:2018lyp}~(red shaded region) are also shown. For $\tan \beta=2$, the reach from PICO-60 is upto about $M_{\widetilde{\chi}^0_1} = 840$~GeV, while the future projection from LZ can probe DM masses upto $1560$~GeV. For $\tan \beta=10$, the reach from PICO-60 is reduced to $230$~GeV and that of LZ to around $350$~GeV.  
 
Thus in the particular simple model adopted in this study the tree-level SD scattering has somewhat better prospects in probing the model parameter space, compared to the one-loop SI scattering rates. However, since the SI and SD rates probe the coupling of the DM particle to different sets of SM particles, both of them are necessary probes of the model, with combined experimental observations leading to a unified picture of the DM-nucleon effective couplings. 

Before concluding, a special remark is in order. The search for missing particles, the potential DM candidates, at high-energy colliders is complementary to the DM direct detection. The charged and neutral dark sector states can be pair-produced in quark-antiquark annihilation via the $s$-channel $W^\pm, Z$-boson exchange in the Drell-Yan process. These states, apart from the lightest neutral DM particle, would decay via electroweak interactions to final states containing $W$ and $Z$ bosons. Thus, multiple leptons and missing transverse momenta are the most promising channels to search for at hadron colliders,  such as the LHC~\cite{Aaboud:2018jiw, Sirunyan:2017lae}, including its luminosity (HL-LHC)~\cite{ ATL-PHYS-PUB-2014-010, CMS-PAS-SUS-14-012}, and possibly energy (HE-LHC) upgrades~\cite{Aboubrahim:2018bil}. For small mass gaps between the charged and neutral dark sector particles, a likely scenario under our consideration, searches for disappearing tracks and displaced vertices are relevant~\cite{Han:2018wus}. For a detailed discussion of the LHC complementarity for DM search near the blind-spot region, we refer the reader to Ref.~\cite{Han:2016qtc}. On the other hand, the situation could be more optimistic if there are relatively light colored states (such as gluinos and squarks in SUSY), that could be  copiously produced at hadron colliders and that could subsequently decay to the DM states resulting in large missing transverse momentum and multiple jets in the final state~\cite{Aaboud:2016zdn}.


\section{Summary}
\label{sec:sum}

In this paper, we studied the NLO electroweak corrections to spin-independent dark matter nucleon scattering, in scenarios where the tree-level predictions for these rates are very small. Such small leading order rates are obtained generically in dark matter models where the DM state results from the mixing of electroweak singlet and doublet states, due to cancellations in DM coupling to the Higgs boson, which is the primary mediator of SI interactions for Majorana fermion WIMPs. A well-known example of these DM blind spots is the case of bino-Higgsino mixed DM in the MSSM. To understand the impact of radiative corrections to DM-nucleon scattering in such a setup, we adopted a simple model for DM with one Majorana fermion singlet, and two electroweak doublets with opposite hypercharge, the neutral components of which mix after electroweak symmetry breaking. This corresponds to the MSSM neutralino sector with all the sfermions, heavy scalars and wino decoupled. 

We evaluated, adopting an on-shell renormalization scheme for the dark matter sector, the set of triangle and box diagrams for the radiative corrections to the DM-quark scalar effective operator, that could directly modify the predictions near the blind spots. We observed that the contribution to the DM-nucleon effective coupling $f_N$ from the triangle diagrams dominates near the tree-level blind spot, as the leading order contribution is vanishingly small in this region. As expected, the one-loop contributions ``unblind'' the tree-level blind spots, as seen in Fig.~\ref{fig:amp}. Away from the blind-spot region, the one-loop electroweak effects are still found to be appreciable. For example, the triangle diagrams considered can shift the tree-level value of $f_N$ by upto $10\%$. We also find that the box diagram contribution can become comparable to the triangles in some parameter regions. There are values of parameters around which both the triangle and the box contributions can also change sign, and therefore have their own blind spots.

Importantly, we always find a new blind spot at the NLO level where the sum of the tree-level and one-loop amplitudes go to zero. This leads to a shifted location for the blind-spot point, the amount of the shift in the values of the doublet mass mixing parameter $M_D$ being almost linearly proportional to the value of the singlet mass $M_S$.  This shift is found to be larger for large values of $\tan \beta$ (the ratio of the Yukawa couplings of the two doublets, $y_1/y_2$) and small values of $y$ (=$\sqrt{y_1^2+y_2^2}$), and can be around $\mathcal{O}(1\%)$. These features are shown in Fig.~\ref{fig:DBD}.

On taking into account the impact of the radiative corrections to SI scattering, the prospects of testing such tree-level blind-spot scenarios in future multi-ton scale liquid Xenon experiments improve considerably. In particular, we find that for smaller values of $\tan \beta$, e.g., $\tan \beta=2$, $\sigma_{\rm SI}$ takes values in the range of about $10^{-47} {~\rm cm}^2$ to $10^{-50} {~\rm cm}^2$, for $M_{\widetilde{\chi}^0_1}$ values in the interval $100 {~\rm GeV} - 2 {~\rm TeV}$, and coupling coefficient $y$ in the range $0.3-1.5$. For this range of couplings, the future projection of the LZ experiment is expected to probe a DM mass upto about $500$ GeV, while the reach can be further extended by the DARWIN experiment upto a DM mass of  $1250$ GeV. On the other hand, for higher values of $\tan \beta$, as seen with $\tan \beta=10$,  the expected cross-section is smaller, with a maximum of around $10^{-49} {~\rm cm}^2$, which may not be accessible to DARWIN. Thus, the small $\tan \beta$ scenario leads to similar $\sigma_{\rm SI}$ as in the case of wino-like real triplet DM, while the intermediate $\tan \beta$ scenario predicts cross-sections similar to the case of Majorana Higgsino-like doublets. These results are presented in Fig.~\ref{fig:SIBD}. On the other hand, as already examined in Ref.~\cite{Han:2016qtc}, the SD scattering cross-sections may be observable in certain SI blind-spot regions. Thus, combined tests of both the SI one-loop predictions and the tree-level SD cross-sections are feasible, thereby probing all the relevant effective operators for DM-nucleon interaction.

With the increasing sensitivity of the dark matter direct detection experiments,  resulting from the construction of bigger and ultra-low noise detectors, it is important to define benchmark targets for these near future multi-ton scale experiments. As we found in this study, higher order electroweak corrections to scenarios with mixed electroweak DM states present one such target, where the tree-level rates can be very small due to the vanishing of relevant DM effective couplings in certain parameter regions. In order to thoroughly probe interesting and well-motivated WIMP scenarios, it is therefore necessary to have theoretical predictions with increased accuracy that could match up to the future expected experimental precision.

\acknowledgments

We would like to thank Ayres Freitas for helpful discussions. This work is supported in part by the U.S.~Department of Energy under grant No.~DE-FG02- 95ER40896 and by the PITT PACC. XW is also supported in part by an Andrew Mellon Predoctoral Fellowship from the School of Arts and Sciences at the University of Pittsburgh.

\begin{appendices}
\section{Details of on-shell renormalization scheme}
\label{renormalization}
The Lagrangian of the DM sector in the mass basis can be written as
\beq
\lag=\overline{\widetilde{\chi}^{+}}\left(\slashed{p}P_L+\slashed{p}P_R-\eta\,M_D\right)\widetilde{\chi}_0^{+} + \frac{1}{2}\overline{\widetilde{\chi}_{i}^{0}}\left(\slashed{p}P_L\delta_{ij}+\slashed{p}P_R\delta_{ij}-\left[\mathbf{U}^\top \mathbf{M_N} \mathbf{U}\right]_{ij}\right)\widetilde{\chi}_{j}^{0},
\eeq
where $\eta$ is a phase factor and $i,j$ are summed over 1 to 3. We specify the on-shell renormalization scheme adopted for the DM sector in the following. 

According to the multiplicative renormalization procedure, we perform the following replacements of the parameters and the fields:
\begin{alignat}{2}
M_S~&\rightarrow~~ M_S+\delta M_S,&\qquad M_D~\rightarrow&~~ M_D+\delta M_D,\\
y_{1}~&\rightarrow~~ y_{1}+\delta y_{1},&\qquad y_{2}~\rightarrow&~~ y_{2}+\delta y_{2},\\
P_L\widetilde{\chi}^+~&\rightarrow~ \left[1 + \frac{1}{2} \delta Z^{L}_{\widetilde{\chi}^+}\right]P_L\widetilde{\chi}^+,\qquad
&P_R\widetilde{\chi}^+~\rightarrow&~ \left[1 + \frac{1}{2} \delta Z^{R}_{\widetilde{\chi}^+}\right]P_R\widetilde{\chi}^+,\\ P_L\widetilde{\chi}^0_i~&\rightarrow~ \left[\mathbb{1} + \frac{1}{2} \delta \mathbf{Z}_{\widetilde{\chi}^0}\right]_{ij}P_L\widetilde{\chi}^0_j,\qquad
&P_R\widetilde{\chi}^0_i~\rightarrow&~ \left[\mathbb{1} + \frac{1}{2} \delta \mathbf{Z}^*_{\widetilde{\chi}^0}\right]_{ij}P_R\widetilde{\chi}^0_j.
\end{alignat}
We note that the transformation matrix $\mathbf{U}$ is not renormalized in our scheme, so that, the mass matrix in the gauge basis $\mathbf{M_N}$ is replaced by
\begin{align}
\mathbf{M_N}~\rightarrow&~\mathbf{M_N}+\delta\mathbf{M_N} = \mathbf{M_N}+\begin{pmatrix}
	\delta M_S & \delta \Delta_2 &  \delta \Delta_1\\
	 \delta \Delta_2 & 0 &  \delta M_D\\
	 \delta \Delta_1 &  \delta M_D & 0\\
     \end{pmatrix}
     \label{ct_matrix},
\end{align}
where $\delta\Delta_{1,2}=\delta(y_{1,2}v/\sqrt{2})$. Then the mass matrix in the mass basis can be expressed as
\beq
\mathbf{M}_{\widetilde{\chi}^0}~\rightarrow~\mathbf{M}_{\widetilde{\chi}^0} + \delta \mathbf{M}_{\widetilde{\chi}^0} = {\rm diag}\left(M_{\widetilde{\chi}^0_1},M_{\widetilde{\chi}^0_2},M_{\widetilde{\chi}^0_3}\right) + \mathbf{U}^\top \delta\mathbf{M_N} \mathbf{U}.
\eeq
In the following, we use $\Sigma$ and $\hat{\Sigma}$ to denote un-renormalized and renormalized self-energies respectively. Decomposing into the following form
\beq
\hat{\Sigma}(p)=\hat{\Sigma}^L(p^2)\slashed{p}P_L+\hat{\Sigma}^R(p^2)\slashed{p}P_R+\hat{\Sigma}^{SL}(p^2)P_L+\hat{\Sigma}^{SR}(p^2)P_R,
\eeq
the renormalized self-energies of the charged and neutral states are given by
\begin{align}
\hat{\Sigma}^L_{\widetilde{\chi}^+}(p^2)&=\Sigma^L_{\widetilde{\chi}^+}(p^2)+\frac{1}{2}( \delta Z^{L}_{\widetilde{\chi}^+}+ \delta Z^{L*}_{\widetilde{\chi}^+}),\\
\hat{\Sigma}^R_{\widetilde{\chi}^+}(p^2)&=\Sigma^R_{\widetilde{\chi}^+}(p^2)+\frac{1}{2}( \delta Z^{R}_{\widetilde{\chi}^+}+ \delta Z^{R*}_{\widetilde{\chi}^+}),\\
\hat{\Sigma}^{SL}_{\widetilde{\chi}^+}(p^2)&=\Sigma^{SL}_{\widetilde{\chi}^+}(p^2)-\frac{1}{2}(M_{\widetilde{\chi}^+} \delta Z^{L}_{\widetilde{\chi}^+}+ \delta Z^{R*}_{\widetilde{\chi}^+}M_{\widetilde{\chi}^+}+2\delta M_{\widetilde{\chi}^+} ),\\
\hat{\Sigma}^{SR}_{\widetilde{\chi}^+}(p^2)&=\Sigma^{SR}_{\widetilde{\chi}^+}(p^2)-\frac{1}{2}(M_{\widetilde{\chi}^+} \delta Z^{R}_{\widetilde{\chi}^+}+ \delta Z^{L*}_{\widetilde{\chi}^+}M_{\widetilde{\chi}^+}+2\delta M^*_{\widetilde{\chi}^+} ),\\
	\left[\hat{\Sigma}^L_{\widetilde{\chi}^0}(p^2)\right]_{ij}&=\left[\Sigma^L_{\widetilde{\chi}^0}(p^2)\right]_{ij}+\frac{1}{2}\left[ \delta \mathbf{Z}_{\widetilde{\chi}^0}+ \delta \mathbf{Z}^\dagger_{\widetilde{\chi}^0}\right]_{ij},\\
	\left[\hat{\Sigma}^R_{\widetilde{\chi}^0}(p^2)\right]_{ij}&=\left[\Sigma^R_{\widetilde{\chi}^0}(p^2)\right]_{ij}+\frac{1}{2}\left[ \delta \mathbf{Z}^*_{\widetilde{\chi}^0}+ \delta \mathbf{Z}^\top_{\widetilde{\chi}^0}\right]_{ij},\\
	\left[\hat{\Sigma}^{SL}_{\widetilde{\chi}^0}(p^2)\right]_{ij}&=\left[\Sigma^{SL}_{\widetilde{\chi}^0}(p^2)\right]_{ij}-\frac{1}{2}\left[\mathbf{M}_{\widetilde{\chi}^0} \delta \mathbf{Z}_{\widetilde{\chi}^0}+ \delta \mathbf{Z}^\top_{\widetilde{\chi}^0}\mathbf{M}_{\widetilde{\chi}^0}+2\delta \mathbf{M}_{\widetilde{\chi}^0}\right]_{ij},\\
	\left[\hat{\Sigma}^{SR}_{\widetilde{\chi}^0}(p^2)\right]_{ij}&=\left[\Sigma^{SR}_{\widetilde{\chi}^0}(p^2)\right]_{ij}-\frac{1}{2}\left[\mathbf{M}_{\widetilde{\chi}^0} \delta \mathbf{Z}^*_{\widetilde{\chi}^0}+ \delta \mathbf{Z}^\dagger_{\widetilde{\chi}^0}\mathbf{M}_{\widetilde{\chi}^0}+2\delta \mathbf{M}^\dagger_{\widetilde{\chi}^0}\right]_{ij}.
\end{align}
We choose the on-shell renormalization scheme by imposing (for $i,j=1,2,3$)
\begin{alignat}{2}
\left.\left[\widetilde{\rm Re}\hat{\Sigma}_{\widetilde{\chi}^+}(p)\right]\widetilde{\chi}^+(p)\right|_{p^2=M_{\widetilde{\chi}^+}^2}&=0,\qquad\lim_{p^2\rightarrow M_{\widetilde{\chi}^+}^2}\frac{1}{\slashed{p}-M_{\widetilde{\chi}^+}}\left[\widetilde{\rm Re}\hat{\Sigma}_{\widetilde{\chi}^+}(p)\right]\widetilde{\chi}^+(p)&=0,\label{os_c}\\
\left.\left[\widetilde{\rm Re}\hat{\Sigma}_{\widetilde{\chi}^0}(p)\right]_{ij}\widetilde{\chi}^0_j(p)\right|_{p^2=M_{\widetilde{\chi}^0_j}^2}&=0,\qquad\lim_{p^2\rightarrow M_{\widetilde{\chi}^0_i}^2}\frac{1}{\slashed{p}-M_{\widetilde{\chi}^0_i}}\left[\widetilde{\rm Re}\hat{\Sigma}_{\widetilde{\chi}^0}(p)\right]_{ii}\widetilde{\chi}^0_i(p)&=0,\label{os_n}
\end{alignat}
where $\widetilde{\rm Re}$ takes only the real part of the loop integrals appearing in the self energies but not of the mixing matrix elements or couplings appearing therein. We further fix the imaginary parts of the wave-function renormalization constants by choosing 
\begin{align}
{\rm Im}\left[\delta Z^L_{\widetilde{\chi}^+}\right] = {\rm Im}\left[\delta Z^R_{\widetilde{\chi}^+}\right] = {\rm Im}\left[\delta \mathbf{Z}_{\widetilde{\chi}^0}\right]_{ii} = 0.\label{os_im}
\end{align}
Thus, Eqs.~(\ref{os_c}$-$\ref{os_im}) yield the counterterms
\begin{align}
\begin{split}
\delta Z^L_{\widetilde{\chi}^+}=&-\Sigma^L_{\widetilde{\chi}^+}(M^2_{\widetilde{\chi}^+})-M^2_{\widetilde{\chi}^+}\left[\Sigma^{L'}_{\widetilde{\chi}^+}(M^2_{\widetilde{\chi}^+})+\Sigma^{R'}_{\widetilde{\chi}^+}(M^2_{\widetilde{\chi}^+})\right]\label{ct_first}\\
&-M_{\widetilde{\chi}^+}\left[\Sigma^{SL'}_{\widetilde{\chi}^+}(M^2_{\widetilde{\chi}^+})+\Sigma^{SR'}_{\widetilde{\chi}^+}(M^2_{\widetilde{\chi}^+})\right],
\end{split}\\
\begin{split}
\delta Z^R_{\widetilde{\chi}^+}=&-\Sigma^R_{\widetilde{\chi}^+}(M^2_{\widetilde{\chi}^+})-M^2_{\widetilde{\chi}^+}\left[\Sigma^{L'}_{\widetilde{\chi}^+}(M^2_{\widetilde{\chi}^+})+\Sigma^{R'}_{\widetilde{\chi}^+}(M^2_{\widetilde{\chi}^+})\right]\\
&-M_{\widetilde{\chi}^+}\left[\Sigma^{SL'}_{\widetilde{\chi}^+}(M^2_{\widetilde{\chi}^+})+\Sigma^{SR'}_{\widetilde{\chi}^+}(M^2_{\widetilde{\chi}^+})\right],
\end{split}\\
\delta M_D =&~\eta^*\delta M_{\widetilde{\chi}^+}=~\frac{\eta^*}{2}M_{\widetilde{\chi}^+}\left[\Sigma^{L}_{\widetilde{\chi}^+}(M^2_{\widetilde{\chi}^+})+\Sigma^{R}_{\widetilde{\chi}^+}(M^2_{\widetilde{\chi}^+})\right]+\eta^*\Sigma^{SL}_{\widetilde{\chi}^+}(M^2_{\widetilde{\chi}^+}),\\
    \begin{split}
	\left[\delta \mathbf{Z}_{\widetilde{\chi}^0}\right]_{ii}=&-\frac{1}{2}\left[\Sigma^L_{\widetilde{\chi}^0}(M^2_{\widetilde{\chi}^0_i})+\Sigma^R_{\widetilde{\chi}^0_i}(M^2_{\widetilde{\chi}^0_i})\right]_{ii} \\
    &-M^2_{\widetilde{\chi}^0_i}\left[\Sigma^{L'}_{\widetilde{\chi}^0}(M^2_{\widetilde{\chi}^0_i})+\Sigma^{R'}_{\widetilde{\chi}^+}(M^2_{\widetilde{\chi}^0_i})\right]_{ii}-M_{\widetilde{\chi}^0_i}\left[\Sigma^{SL'}_{\widetilde{\chi}^0}(M^2_{\widetilde{\chi}^0_i})+\Sigma^{SR'}_{\widetilde{\chi}^0}(M^2_{\widetilde{\chi}^0_i})\right]_{ii},
    \end{split}\\
    \begin{split}
	\left[\delta \mathbf{Z}_{\widetilde{\chi}^0}\right]_{ij}=&~\frac{2}{M^2_{\widetilde{\chi}^0_i}-M^2_{\widetilde{\chi}^0_j}}\left[M^2_{\widetilde{\chi}^0_j}\Sigma^L_{\widetilde{\chi}^0}(M^2_{\widetilde{\chi}^0_j})+M_{\widetilde{\chi}^0_i}M_{\widetilde{\chi}^0_j}\Sigma^R_{\widetilde{\chi}^0}(M^2_{\widetilde{\chi}^0_j})+M_{\widetilde{\chi}^0_i}\Sigma^{SL}_{\widetilde{\chi}^0}(M^2_{\widetilde{\chi}^0_j})\right.\\
    &\left.+M_{\widetilde{\chi}^0_j}\Sigma^{SR}_{\widetilde{\chi}^0}(M^2_{\widetilde{\chi}^0_j})-M_{\widetilde{\chi}^0_i}\delta \mathbf{M}_{\widetilde{\chi}^0}-M_{\widetilde{\chi}^0_j}\delta \mathbf{M}^\dagger_{\widetilde{\chi}^0}\right]_{ij},\quad {\rm for~}i\neq j,
    \end{split}\\
	\left[\delta \mathbf{M}_{\widetilde{\chi}^0}\right]_{ii}=&~\frac{1}{2}M_{\widetilde{\chi}^0_i}\left[\Sigma^{L}_{\widetilde{\chi}^0}(M^2_{\widetilde{\chi}^0_i})+\Sigma^{R}_{\widetilde{\chi}^0}(M^2_{\widetilde{\chi}^0_i})\right]_{ii}+\left[\Sigma^{SL}_{\widetilde{\chi}^0}(M^2_{\widetilde{\chi}^0_i})\right]_{ii},\label{ct_last}	
\end{align}
where $\Sigma'(p^2)$ is the derivative of the self-energy $\Sigma'(p^2)=\partial\Sigma(p^2)/\partial p^2$. All the un-renormalized self-energies $\Sigma$ in Eqs.~(\ref{ct_first}$-$\ref{ct_last}) should be understood as $\widetilde{\rm Re}\Sigma$.
The counterterms $\delta\Delta_1$, $\delta\Delta_2$ and $\delta M_S$ in Eq.~(\ref{ct_matrix}) are then fixed by solving the equations
\beq
\left[\delta\mathbf{M}_{\widetilde{\chi}^0}\right]_{ii} = \left[\mathbf{U}^\top\delta\mathbf{M_N}\mathbf{U}\right]_{ii}.
\eeq
The relevant counterterms to the DM-Higgs coupling can be expressed as
\beq
\begin{split}
\delta\Gamma^{\rm ct}\left(\overline{\widetilde{\chi}_{1}^0},\widetilde{\chi}_{1}^0,h\right)=&~~~\left[\delta \mathbf{S}^L
+\frac{1}{2}\delta \mathbf{Z}_{\widetilde{\chi}^0}^\top \mathbf{S}^L+\frac{1}{2}\mathbf{S}^L\delta \mathbf{Z}_{\widetilde{\chi}^0}+\frac{1}{2}\mathbf{S}^L\delta Z_h\right]_{11} P_L \\
&+\left[\delta \mathbf{S}^R+\frac{1}{2}\delta \mathbf{Z}_{\widetilde{\chi}^0}^\dagger \mathbf{S}^R+\frac{1}{2}\mathbf{S}^R\delta \mathbf{Z}_{\widetilde{\chi}^0}^*+\frac{1}{2}\mathbf{S}^R\delta Z_h\right]_{11} P_R.
\end{split}
\eeq
where
\begin{align}
\left[\mathbf{S}^L\right]_{ii}&=-\frac{y_{1}}{\sqrt{2}}U_{3i}U_{1i}-\frac{y_{2}}{\sqrt{2}}U_{2i}U_{1i},\\
\left[\mathbf{S}^L\right]_{ij}&=-\frac{y_{1}}{\sqrt{2}}U_{3i}U_{1j}-\frac{y_{2}}{\sqrt{2}}U_{2i}U_{1j} + (i\leftrightarrow j),\quad {\rm for~}i\neq j,\\
\mathbf{S}_{R}&=\mathbf{S}_{L}^{\dagger}.
\end{align}
The counterterms $\delta y_1$ and $\delta y_2$ are related to $\delta \Delta_1$ and $\delta \Delta_2$ through the relations
\begin{align}
\delta y_1 &= \sqrt{2} \frac{\delta \Delta_1}{v}-y_1\frac{\delta v}{v},\qquad\delta y_2 = \sqrt{2} \frac{\delta \Delta_2}{v}-y_2\frac{\delta v}{v},
\end{align}
with $\delta v$ and $\delta Z_h$ calculated in the on-shell scheme following the conventions in Ref.~\cite{Denner:1991kt}.

\section{DM-nucleon scattering: computational framework}
\label{DD}
In this Appendix, we briefly review the formalism adopted for computing the DM-nucleon scattering cross-sections~\cite{Jungman:1995df}, and the values of the relevant nuclear matrix elements used. The effective interactions of a non-relativistic Majorana WIMP $X$ with light quarks and gluons are given as
\beq
\lag_{eff}=\sum_{q=u,d,s} \left(d_q\overline{X}\gamma^\mu\gamma^5X\bar{q}\gamma_\mu\gamma^5q+f_qm_q\overline{X}X\bar{q}q \right)+f_G\overline{X}X\frac{\alpha_s}{\pi}G^a_{\mu\nu}G^{a\mu\nu},
\eeq
where, $G^a_{\mu\nu}$ is the gluon field strength tensor and $\alpha_S$ is the strong coupling constant. Here, the operator involving axial-vector currents of the DM and the quark fields leads to spin-dependent interactions, while the other two operator structures lead to spin-independent scattering with nuclei.

To begin with, we define the matrix element (ME) of the scalar operator $\bar{q}q$ between nucleon states $N$ (where $N$ is either a proton or a neutron) as follows:
\beq
\langle N | m_q \bar{q}q | N \rangle \equiv f_{Tq}^N m_N.
\eeq
The corresponding ME of the gluon operator can be obtained by using the trace of the energy momentum tensor $T^\mu_\mu$, which is given by
\beq
T^\mu_\mu=\sum_{q=u,d,s}m_q\bar{q}q+\sum_{Q=b,c,t}m_Q\bar{Q}Q-\frac{7\alpha_s}{8\pi}GG. 
\eeq
Here, we have used the shorthand $GG$ to stand for $G^a_{\mu\nu}G^{a\mu\nu}$. Utilizing the fact that 
\beq
\langle N|T^\mu_\mu|N\rangle \equiv m_N,
\eeq
where, $m_N$ is the nucleon mass, and by integrating out the heavy quarks using
\beq
\langle N|m_Q\bar{Q}Q|N\rangle = \langle N|-\frac{\alpha_s}{12\pi}GG|N\rangle,
\eeq
we obtain the ME of the gluon operator 
\beq
\langle N|\frac{\alpha_s}{\pi}GG|N\rangle =-\frac{8}{9} m_N f^N_{TG}.
\eeq
Here,  $f^N_{TG}$ is related to $f^N_{Tq}$ as
\beq
f^N_{TG}\equiv 1-\sum_{q=u,d,s}f^N_{Tq}.
\eeq

Similarly, the nucleon ME of the axial-vector quark current is defined as 
\beq
\langle N|\bar{q}\gamma_\mu\gamma^5q|N\rangle \equiv 2 s_\mu \Delta q_N,
\eeq
where $s_\mu$ is the nucleon spin. Combining these results, the effective interaction of Majorana WIMPs with nucleons is given by
\beq
\lag_{eff}=\sum_{N=n,p} \left (f_N\overline{X}X\overline{N}N+a_N\overline{X}\gamma^\mu\gamma^5X\overline{N}\gamma_\mu\gamma^5N \right),
\eeq
with the Wilson co-efficients,
\beq
\begin{split}
	f_N/m_N= \sum_{q=u,d,s} f_q f^N_{Tq}-\frac{8}{9}f_G f^N_{TG}\quad 
	{\rm and~~}a_N=\sum_{q=u,d,s}d_q\Delta q_N.
\end{split}
\eeq
For our computations, we adopt the following values of the nuclear matrix elements for proton: $f_{Tu}^p=0.0153$, $f_{Td}^p=0.0191$, and $f_{Ts}^p=0.0447$, where we have used the lattice results for the strange quark content of the nucleon~\cite{Alvarez-Ruso:2013fza,Junnarkar:2013ac,Belanger:2013oya}. For spin-dependent scattering, we use the following inputs: $\Delta u_p = 0.842$, $\Delta d_p= -0.427$, and $\Delta s_p= -0.085$~\cite{Belanger:2013oya}. 

\section{Mapping the singlet-doublet model to MSSM}
\label{SUSY}
The analysis presented in Sec.~\ref{sec:SD} can be translated to the neutralino sector in the minimal supersymmetric standard model (MSSM), with the wino state decoupled. In such a scenario,  the neutralino mass matrix in the basis $(\widetilde{B},\widetilde{H}^0_d,\widetilde{H}^0_u)$ is given by
\beq
M_N=\begin{pmatrix}
	M_1  & -M_Z s_W \cos\beta & M_Z s_W \sin\beta \\
	-M_Z s_W \cos\beta  & 0 & -\mu\\
	M_Z s_W \sin\beta & -\mu & 0\\
\end{pmatrix}
.
\eeq
The phenomenology of tree-level spin-independent DM-quark interactions is then similar to what we obtained for the singlet-doublet model, with the following mapping between the couplings,
\beq
\begin{gathered}
	y\rightarrow -\sqrt{2}\frac{M_Z s_W}{v}.
\end{gathered}
\eeq
The singlet and doublet fermion mass parameters $M_S$ and $M_D$ are replaced by the bino and Higgsino mass parameters, $M_1$ and $\mu$, respectively.
The coupling of the lighter Higgs boson state to the lightest neutralino is then given by
\beq
g^{0}_{h\chi_1\chi_1} \simeq \frac{e M_Z \tan\theta_W}{\mu^2-M_1^2} \left(M_1+\mu\sin(2\beta) \right).
\eeq
In the MSSM, we also have the following DM coupling to the heavier CP-even Higgs boson
\beq
g^{0}_{H\chi_1\chi_1}\simeq -\frac{e M_Z \tan\theta_W}{\mu^2-M_1^2}\mu\cos(2\beta).
\eeq
Combining with the Higgs-quark Yukawa couplings and taking the alignment limit, at the leading order the DM-quark scalar effective couplings are then obtained to be
\beq
\begin{split}
	f_u=-\frac{g^{0}_{h\chi_1\chi_1}}{vm_h^2}+\frac{g^{0}_{H\chi_1\chi_1}}{vm_H^2}\cot\beta,\quad 
	f_d=-\frac{g^{0}_{h\chi_1\chi_1}}{vm_h^2}-\frac{g^{0}_{H\chi_1\chi_1}}{vm_H^2}\tan\beta.
\end{split}
\eeq
In the scenario with the heavy Higgs decoupled, we can now obtain the SI blind-spot condition for MSSM:
\beq
M_1+\mu\sin(2\beta)=0,
\eeq
with ${\rm sgn} \left(M_1/\mu \right)=-1$.

\end{appendices}

\bibliographystyle{JHEP}
\bibliography{ref}

\end{document}